\documentclass[acmsmall, screen]{acmart}
\AtBeginDocument{%
  }

\setcopyright{acmlicensed}
\copyrightyear{2026}
\acmYear{2026}
\acmDOI{10.1145/3822173}
\acmJournal{TACO}
\acmISBN{978-1-4503-XXXX-X/18/06}

\usepackage[most]{tcolorbox}
\usepackage{enumitem}

\usepackage{todonotes}
\usepackage{tikz}
\usetikzlibrary{tikzmark}

\usepackage{setspace}

\usepackage{xcolor}
 \definecolor{darkspringgreen}{rgb}{0.09, 0.45, 0.27}
\definecolor{denim}{rgb}{0.08, 0.38, 0.74}
\definecolor{darkolivegreen}{rgb}{0.33, 0.42, 0.18}
\definecolor{tangerine}{rgb}{0.95, 0.52, 0.0}
\definecolor{mahogany}{rgb}{0.75, 0.25, 0.0}
\definecolor{pred}{rgb}{0.7843, 0.0039, 0.3137}

\definecolor{seagreen}{rgb}{0.18, 0.55, 0.34}

\definecolor{darkpink}{rgb}{0.88, 0.28, 0.54}
\definecolor{forestgreen}{rgb}{0.0, 0.27, 0.13}
\definecolor{amber}{rgb}{1.0, 0.49, 0.0}

\newcommand\gf[1]{{\color{black}{#1}}}

\newcommand{\sect}[1]{{§#1}\xspace} %
\newcommand{\head}[1]{{\noindent\textbf{#1.}\xspace}} %
\newcommand{\fig}[1]{{Fig.~#1}\xspace} %

\usepackage{tikz}

\newcommand{\squishlist}{
 \begin{list}{$\circ$}
  { \setlength{\itemsep}{0pt}
     \setlength{\parsep}{0pt}
     \setlength{\topsep}{3pt}
     \setlength{\partopsep}{0pt}
     \setlength{\leftmargin}{1em}
     \setlength{\labelwidth}{1em}
     \setlength{\labelsep}{0.5em} } }

\newcommand{\squishend}{
  \end{list}  }

\usepackage[compact]{titlesec}

\titlespacing\section{0pt}{4pt plus 1pt minus 1pt}{4pt plus 1pt minus 2pt}
\titlespacing\subsection{0pt}{4pt plus 1pt minus 1pt}{5pt plus 1pt minus 2pt}
\titlespacing\subsubsection{0pt}{2pt plus 1pt minus 1pt}{3pt plus 1pt minus 2pt}

\usepackage{setspace}

\newcommand*\circled[1]{\tikz[baseline=(char.base)]{\node[shape=circle,fill,inner sep=0.5pt] (char) {\textcolor{white}{#1}};}}
\definecolor{denim}{rgb}{0.08, 0.38, 0.74}
\definecolor{azure(colorwheel)}{rgb}{0.0, 0.5, 1.0}
\definecolor{greenp}{rgb}{0.0, 0.65, 0.50}
\definecolor{peach}{rgb}{0.97, 0.51, 0.47}
\definecolor{darkmagenta}{rgb}{0.55, 0.0, 0.55}
\definecolor{royalblue(web)}{rgb}{0.25, 0.41, 0.88}
\definecolor{ao(english)}{rgb}{0.0, 0.5, 0.0}
\definecolor{ForestGreen}{RGB}{34,139,34}

\newcommand{\uop}{{\textmu}op\xspace}
\newcommand{\uops}{{\textmu}ops\xspace}
\newcommand{\proposal}{RevaMp3D\xspace}

\usepackage[bottom]{footmisc}
\usepackage{dblfloatfix}
\usepackage{array}
 \usepackage{booktabs}

\definecolor{coolblack}{rgb}{0.0, 0.35, 0.45}
\definecolor{mahogany}{rgb}{0.75, 0.25, 0.0}

\newcommand{\js}[1]{{\color{pred}{#1}}}
\newcommand{\mrev}[1]{{\color{black}{#1}}}

 \newcommand{\rev}[1]{{\color{black}{#1}}}

\newcommand{\romnum}[1]{{(\textit{#1})}}
\definecolor{ashgrey}{rgb}{0.7, 0.75, 0.71}
\definecolor{grayy}{rgb}{0.89, 0.89, 0.89}
\newcommand\new[1]{{\color{black}{#1}}}
\newcommand{\grm}[1]{{\color{black}{#1}}}

\newcommand{\nmicro}[1]{{\color{black}{#1}}}
\newcommand\revmid[1]{\todo[linecolor=magenta,backgroundcolor=magenta!15,bordercolor=magenta]{\textcolor{black}{\textbf{#1}}}}
\renewcommand\revmid[1]{}

\newcommand{\mrevmic}[1]{{\color{black}{#1}}}

\definecolor{lightblue}{rgb}{0.3,0.5,1.0}

\newcommand{\nmg}[1]{{\color{blue}{#1}}}
\newcommand{\omc}[1]{{\color{orange}{#1}}}
\newcommand{\appmove}[1]{{\color{magenta}{#1}}}

\renewcommand{\js}[1]{{\color{black}{#1}}}
\renewcommand{\appmove}[1]{{\color{black}{#1}}}
\renewcommand{\nmg}[1]{{\color{black}{#1}}}
\renewcommand{\omc}[1]{{\color{black}{#1}}}

\definecolor{cornflowerblue}{rgb}{0.39, 0.58, 0.93}
\newcommand{\taco}[1]{{\color{black}{#1}}}

\makeatletter
\g@addto@macro{\normalsize}{%
  \setlength{\abovedisplayskip}{2pt plus 1pt minus 1pt}
  \setlength{\belowdisplayskip}{2pt plus 1pt minus 1pt}
  \setlength{\abovedisplayshortskip}{0pt}
  \setlength{\belowdisplayshortskip}{0pt}
  \setlength{\intextsep}{2pt plus 1pt minus 1pt}
  \setlength{\textfloatsep}{3pt plus 1pt minus 1pt}
  \setlength{\dbltextfloatsep}{3pt plus 1pt minus 1pt}
  \setlength{\skip\footins}{4pt plus 1pt minus 1pt}}
\makeatother

\newcommand{\omf}[1]{{\color{black}{#1}}}
\usepackage{xargs}

\newcommand\revida[1]{\todo[noline,linecolor=blue,backgroundcolor=blue!25,bordercolor=blue]{\textcolor{white}{\textbf{#1}}}}

\definecolor{ao(english)}{rgb}{0.0, 0.5, 0.0}

\newcommand\revidb[1]{\todo[noline,linecolor=ao(english),backgroundcolor=ao(english)!25,bordercolor=ao(english)]{\textcolor{white}{\textbf{#1}}}}

\definecolor{purple(html/css)}{rgb}{0.53, 0.0, 0.69}
\definecolor{violet(ryb)}{rgb}{0.53, 0.0, 0.69}
\definecolor{violet}{rgb}{0.56, 0.0, 1.0}
\definecolor{darkviolet}{rgb}{0.58, 0.0, 0.83}
\definecolor{chestnut}{rgb}{0.545, 0.118, 0.247}

\newcommand\revidc[1]{\todo[noline,linecolor=chestnut,backgroundcolor=chestnut!25,bordercolor=chestnut]{\textcolor{white}{\textbf{#1}}}}

\definecolor{coolblack}{rgb}{0.0, 0.18, 0.39}
\definecolor{coolerblack}{rgb}{0.25, 0.385, 0.543}
\newcommand{\revtaco}[1]{{\color{black}{#1}}}

\newcommand{\ch}[2]{{\sethlcolor{#1}\hl{#2}}}

\definecolor{lavendergray}{rgb}{0.77, 0.76, 0.82}
\usepackage{wrapfig}

\definecolor{lightyellow}{rgb}{0.980, 0.956, 0.623}

\usepackage[framemethod=tikz]{mdframed}

\newcommand{\boxbegin} {
	\begin{tcolorbox}[enhanced, frame hidden, colback=gray!50, breakable]
}

\newcommand{\boxend} {
	\end{tcolorbox}
}

\newcommand{\yboxbegin} {
	\begin{tcolorbox}[breakable, enhanced, frame hidden, colback=yellow!50]
}

\newcommand{\yboxend} {
	\end{tcolorbox}
}

\mdfdefinestyle{graybox}{
    splittopskip=0,%
    splitbottomskip=0,%
    frametitleaboveskip=0,
    frametitlebelowskip=0,
    skipabove=0,%
    skipbelow=0,%
    leftmargin=0,%
    rightmargin=0,%
    innertopmargin=0mm,%
    innerbottommargin=0mm,%
    roundcorner=1mm,%
    backgroundcolor=lightyellow,
    hidealllines=true}
    
\mdfdefinestyle{graybox2}{
    splittopskip=0,%
    splitbottomskip=0,%
    frametitleaboveskip=0,
    frametitlebelowskip=0,
    skipabove=0,%
    skipbelow=0,%
    leftmargin=0,%
    rightmargin=0,%
    innertopmargin=2mm,%
    innerbottommargin=2mm,%
    roundcorner=2mm,%
    backgroundcolor=lightyellow,
    hidealllines=true}

\newcommand{\bboxbegin}{
    \begin{mdframed}[style=graybox]
}

\newcommand{\bboxend}{
    \end{mdframed}
}

\newcommand{\yyboxbegin}{
    \begin{mdframed}[style=graybox2]
}
\newcommand{\yyboxbegintitle}[1]{
    \begin{mdframed}[style=graybox2,frametitle=#1]
}

\newcommand{\yyboxend}{
    \end{mdframed}
}

\definecolor{almond}{rgb}{0.94, 0.87, 0.8}
\definecolor{antiquebrass}{rgb}{0.8, 0.58, 0.46}

\renewcommand\revidc[1]{}
\renewcommand\revidb[1]{}
\renewcommand\revida[1]{}

\newcommand{\omct}[1]{{\color{magenta}{#1}}}
\renewcommand{\omct}[1]{{\color{black}{#1}}}
  \usepackage{verbatim}
\usepackage{soul}
 \usepackage{framed}
\usepackage[utf8]{inputenc}

 \usepackage{listings}
 \usepackage{xcolor}
 \usepackage{multirow}
\usepackage{booktabs}
 \usepackage[most]{tcolorbox}
 \usepackage{xspace}
\usepackage[caption=false]{subfig}
\usepackage[bottom]{footmisc}
\usepackage{dblfloatfix}
\usepackage{array}
\usepackage{clipboard}

\begin{document}

\title[RevaMp3D]{RevaMp3D: Architecting the Processor Core and Cache Hierarchy \\for Systems with Monolithically-Integrated Logic and Memory}

\author{Nika Mansouri Ghiasi}
\orcid{0000-0002-0833-0042}
\author{Mohammad Sadrosadati}
\orcid{0000-0002-4029-0175}
\author{Geraldo F. Oliveira}
\orcid{0000-0003-1557-4819}
\author{Konstantinos Kanellopoulos}
\orcid{0000-0002-2375-7490}
\affiliation{%
  \institution{ETH Zurich}
  \country{Switzerland}}

\author{Rachata Ausavarungnirun}
\orcid{0000-0002-1459-0852}
\affiliation{
\institution{MangoBoost}
\country{USA}}

\author{Juan G\'omez Luna}
\orcid{0000-0002-6514-1571}
\author{João Ferreira}
\orcid{0000-0003-3855-3906}
\author{Jeremie S. Kim}
\orcid{0000-0001-6153-9008}
\affiliation{
\institution{ETH Zurich}
\country{Switzerland}
}

\author{Christina Giannoula}
\orcid{0000-0003-0162-4547}
\affiliation{
\institution{University of Toronto and ETH Zurich}
\country{Canada and Switzerland}
}

\author{Nandita Vijaykumar}
\orcid{0000-0003-3315-9336}
\affiliation{
\institution{University of Toronto}
\country{Canada}
}

\author{Jisung Park}
\orcid{0000-0002-1826-9003}
\affiliation{
\institution{POSTECH}
\country{South Korea}
}

\author{Onur Mutlu}
\orcid{0000-0002-0075-2312}
\affiliation{
\institution{ETH Zurich}
\country{Switzerland}
}

\renewcommand{\shortauthors}{Mansouri Ghiasi et al.}

\begin{abstract}
   Recent nano-technological advances enable the \emph{Monolithic 3D (M3D)} integration of multiple memory and logic layers \emph{in a single chip}. Such integration enables 
high-bandwidth connections between layers, \omc{which significantly alleviates main memory bottlenecks.} 
We show for a variety of workloads, on a state-of-the-art M3D\omf{-based} system, that the performance and energy bottlenecks shift from main memory to the \nmg{processor} core and cache hierarchy.
Therefore, to effectively utilize the chip's area, given the applications' shifted requirements, there is a need to revisit current processor core and cache hierarchy designs that have been \nmg{conventionally} tailored to tackle the memory bottleneck. 
Based on the insights from our 
design space exploration, we propose \textbf{\proposal}, which introduces five key changes to the state-of-the-art M3D-based system. 
First, we propose removing the shared last-level cache, based on our observation that doing so \taco{achieves speedups on par with, or even exceeding, speedups achieved by increasing its size or reducing its latency, \omf{across all our workloads}.}
Second, \taco{since our analysis shows that improving L1 cache latency has a large impact on performance in M3D}, we reduce L1 cache latency by \omf{leveraging} an M3D layout that reduces wire lengths. Third, we leverage the area reclaimed from removing large caches to widen and scale up various structures in the processor pipeline \omf{that enable greater instruction-level parallelism}. 
To avoid latency penalties from these larger structures, we \omf{leverage} M3D layouts that keep their wire lengths short. 
Fourth, to facilitate high \omf{thread-level} parallelism, we propose a \omf{new} fine-grained synchronization technique, using M3D’s dense inter-layer connectivity.
Fifth, \taco{we leverage the M3D main memory to mitigate the performance and energy bottlenecks of the processor core. To this end, we \omf{propose a processor frontend design that} memoizes} the repetitive \nmg{fetched, decoded, and reordered instructions, \omf{stores them} in main memory at low cost,} and turns off \nmg{the} relevant parts of the core when possible. \omf{The high-bandwidth, energy-efficient M3D memory enables storing and loading the memoized instructions efficiently, eliminating the need for large SRAM for storing the instructions.}  
Our evaluation using a wide range of \omf{20} real-world workloads and \omf{7 multi-programmed mixes} shows that
\proposal provides \omf{1.2$\times$--2.9$\times$} speedup and \omf{1.2$\times$--1.4$\times$} energy reduction, while achieving \nmg{12.3}\% smaller area, compared to the \nmg{state-of-the-art} M3D\omf{-based} system. 
\omct{Compared to the state-of-the-art 3D and 2D systems, \proposal provides 4.96$\times$ and 7.14$\times$ average speedup, respectively.}
\omf{We also analyze the impact of \proposal's design decisions for M3D systems with various main memory latency values since latency can vary depending on the design decisions made to meet certain requirements of the target system. This analysis facilitates making the appropriate
design decisions based on 
latency, thereby benefiting a wide range of workloads.}
\vspace{-0.5em}

\end{abstract}

\ccsdesc[500]{Computer systems organization~Architectures}
\ccsdesc[500]{Hardware~Analysis and design of emerging devices and systems}

\keywords{Monolithic 3D Integration, Memory Bottlenecks, Processor Core, Cache Hierarchy, Memory Systems, 3D Stacking}

\maketitle

\section{Introduction}
\label{sec:introduction}

Recent nano-technological advances enable the \emph{Monolithic 3D (M3D)} integration of multiple logic and memory layers on a single chip~\cite{srimani2023n3xt,shulaker2013carbon, walden2021monolithically,ghosh2024monolithic,shulaker2014monolithic, jagasivamani2020tileable,shulaker2017three, aly2015energy, aly2018n3xt,jagasivamani2019analyzing,cheng2022emerging}. M3D integration
\js{is a promising technology to \taco{alleviate} the memory bottleneck that significantly limits the performance and energy efficiency of data-intensive applications in conventional systems}.
M3D \taco{integration} enables \nmg{significantly} higher main memory bandwidth compared to other 3D integration technologies, such as \js{Through Silicon Via (TSV)}-based~\cite{hbmA100, hbm3skhynix, aly2015energy, gopireddy2019designing} or \taco{hybrid bonding-based}~\cite{kim2016ultra,niu2022184qps,fujun2020stacked,arnaud2020three,hu2021development,nigussie2021design} integrations. This is due to M3D's much higher density of \nmg{Inter-Layer Vias (ILVs)} that connect \js{different} device layers.
\taco{M3D integration can also lead to lower main memory latency by exploiting the fine-grained ILVs to efficiently enable \romnum{i}~a large number of memory channels, each connected to relatively small memory arrays with shorter wire lengths, \romnum{ii}~fine-grained 3D layout of the memory array to further shorten the wires, and \romnum{iii}~low-latency interconnection\omf{s} between logic and memory layers on the same chip.}

Due to its benefits, M3D \taco{integration} has received substantial interest in both academia (e.g.,~\cite{srimani2023n3xt,shulaker2013carbon,ghosh2024monolithic, walden2021monolithically,shulaker2014monolithic, jagasivamani2020tileable,shulaker2017three, aly2015energy, aly2018n3xt,jagasivamani2019analyzing,cheng2022emerging,du2024monolithic,lee2025heterogeneous,wang2025hisim,sedaghatgoo2024arman,chang2024design,pentapati2024heterogeneous,kim2023ppa}) and industry (e.g.,~\cite{skywaterm3d,liao2023complementary,subramanian2020first,kim2024growth,athena2024first}). 
Several prior works propose improvements for M3D-based systems, \nmg{mainly focusing} on three aspects.
First, at the device and circuit level\omf{s}, \omf{various} works (e.g., \cite{ku2016physical, liu2012design, panth2014design, chang2017design,samal2014fast, shukla2019overview, lee2018performance, samal2014full, brocard2017transistor, samal2016monolithic, nayak2015power, jagasivamani2019analyzing, walden2021monolithically, jagasivamani2019design, zokaee2019magma, murali2020heterogeneous, chatterjee2021power, murali2021heterogeneous}) propose improvements to the logic and memory components and their integration in M3D substrates.
Second, at the architecture level, some works (e.g., \cite{stow2019efficient,joardar20213d++,gopireddy2019designing, jagasivamani2018memory, srinivasa2018monolithic,srinivasa1,srinivasa2,kong2017architecting}) 
\omf{demonstrate the impact of the M3D integration of logic layers on  the performance and energy efficiency of logic components, such as SRAM structures in processor cores and networks on chips.} 
Other works (e.g., \cite{jagasivamani2018memory,jagasivamani2019analyzing}) explore the memory system challenges in M3D (e.g., designing main memory array structures and controllers).
Third, at the application level, various works (e.g., \cite{felfldate2020,yu2020spring, ko2020efficient, huang2019ragra, hanif2020resistive, joardar20213d++, wu2018brain, wu2018hyperdimensional}) propose application-specific M3D accelerators.

While prior works propose various \new{improvements} for M3D-based systems, no prior work \omf{re-examines}  the processor core and cache hierarchy designs based on the requirements of real-world applications on systems with monolithically-integrated logic and memory.
\omf{Therefore, to effectively leverage M3D integration's benefits}, it is important to understand its implications on the performance and energy bottlenecks of real-world applications. These implications can then guide  \omf{re-examining processor core and cache hierarchy} designs.
Based on our experimental analysis of a variety of workloads \omf{simulated} on a state-of-the-art M3D-based system~\cite{aly2018n3xt, shulaker2014monolithic, shulaker2017three, aly2015energy}, we make two key observations. First, compared to the 2D and 3D systems,  performance bottlenecks \taco{in the M3D-based system} \emph{shift} \omf{substantially} from the main memory to the processor core and caches \nmg{(even for the memory-bound workloads)}. Second, the processor core becomes the primary source of energy consumption \nmg{due to the \omf{more} energy-efficient M3D main memory (see \sect{\ref{sec:background}})}.

\textbf{Our goal} in this work is to design the core and cache hierarchy given the shifted bottlenecks in M3D \taco{systems}.
To this end, we take two steps. First, we explore the implications of \taco{shifted} bottlenecks on \omf{the processor core and cache hierarchy} design. Second, \new{based on these implications} and by leveraging the new opportunities of M3D \taco{integration}, we design a new M3D-based system.

\head{Implications of M3D \taco{\omf{I}ntegration} on \omc{Cache Hierarchy Design}}
Through \omf{a} detailed analysis of the \emph{depth}, \emph{size}, and \emph{latency} of the cache hierarchy \omf{in an M3D-based system}, we make three new observations. 
First, all \omf{evaluated} workloads perform better or equally well in an M3D-based system with a shallow one-level cache hierarchy \omf{(i.e., with L1 caches)} compared to an M3D-based system with \nmg{a deeper and larger cache hierarchy}
(\sect{\ref{sec:cache-depth}}). 
Second, although M3D integration enables \omf{reducing cache latency}, the benefit of removing \nmg{the deeper cache levels} is still higher than or comparable to increasing the size\omf{s} (\sect{\ref{sec:cache-size})} and/or improving the latenc\omf{ies of the deeper cache levels} (\sect{\ref{sec:cache-lat}}). 
Third, \omf{in contrast} to the deeper levels of the cache hierarchy, L1 caches \omf{in an M3D-based system} are still critical for performance, and improving their latency improves performance for a wide range of workloads (\sect{\ref{sec:cache-lat}}).

\head{Implications of M3D \taco{Integration} on \omc{Processor Core Design}}
We perform a detailed analysis of different units in all stages of \omf{an out-of-order superscalar} processor core pipeline.
\taco{We highlight three new observations, demonstrating the most important factors impacting performance.}
First, larger pipeline widths improve M3D performance since they increase the number of in-flight requests, which can be effectively served given M3D's high memory bandwidth (\sect{\ref{sec:core-wide}}). 
Second, \nmg{the performance impact of pipeline frontend \omf{(e.g., fetch, decode, and branch prediction units)} increases significantly} (\sect{\ref{sec:arch-spec}}), as M3D shifts the bottleneck away from memory, increasing the fraction of execution time dominated by these \omf{units}. \nmg{Third, while high-bandwidth main memory enables running many threads concurrently, the overhead \omf{of synchronization between threads in multi-threaded workloads} limits performance (\sect{\ref{sec:sync-opt}})}.

\head{Re-Architecting M3D}
\omf{Considering the implications of M3D integration on cache hierarchy and processor core design}, we propose an optimized M3D-based system, \emph{\proposal}, 
\omc{\omf{based on} five \omf{key} design decisions.
First, we remove the shared last-level cache (L2 in our M3D baseline) \omf{from the cache hierarchy}. 
Second, we reduce L1 cache latency by \taco{devising a 3D layout for the SRAM array, in which the array is divided across two logic layers to reduce its overall wire lengths and latency.}
Third, we leverage the area reclaimed from removing \omf{the} large cache to widen and scale up various structures in the processor pipeline \omf{that enable greater instruction-level parallelism (e.g., reorder buffer and execution units)}.
This design accommodates more in-flight requests, which can be efficiently served due to the high M3D memory bandwidth. \omf{We} avoid latency penalties from these larger structures \omf{by leveraging} M3D layout\omf{s to} keep their wire lengths short.
Both the second and third \omf{design decisions} take advantage of the dense ILVs that enable efficient vertical layout of the processor structures~\cite{stow2019efficient,joardar20213d++,gopireddy2019designing,srinivasa2018monolithic,srinivasa1,srinivasa2,kong2017architecting}. 
Fourth, we \omf{introduce a new} fine-grained register file-level synchronization technique, \omf{which} eliminate\omf{s} the additional latency in the memory hierarchy for synchronization.
This technique is realized by using the thin ILVs to efficiently increase the register file bandwidth to support both the \omf{regular operand accesses} and the synchronization-related accesses.
Fifth, \omf{we leverage the M3D main memory to mitigate the performance and energy bottlenecks of the processor core. To this end, we propose a processor frontend design that memoizes the repetitive (i.e., in a loop) fetched, decoded, and reordered instructions, stores them in main memory at low cost, and turns off the relevant parts of the processor core when possible. The high-bandwidth, energy-efficient M3D main memory enables storing and loading the memoized instructions efficiently, \omct{thereby} eliminating}  the area overhead of prior memoization techniques that require a large SRAM capacity~\cite{talpes2005execution,solomon2003micro}.}
This \omf{design} alleviates \romnum{i}~energy bottlenecks when repeated instructions execute, and \romnum{ii}~branch mis\omf{prediction} overhead since a large fraction of instructions do not need to be fetched, decoded, and re-ordered on branch mis\omf{prediction}.

\nmg{\omc{\head{Key Results}} We evaluate \proposal using a \omf{heavily modified} cycle-accurate architectural
simulator (ZSim~\cite{sanchez2013zsim}) to flexibly explore various designs
and faithfully model the M3D-based systems. 
\omf{We} experimentally calibrate the parameters for the M3D baseline \omf{based on} the fabricated device models, circuit analysis, interconnect model, and physical design tools presented in ~\cite{aly2018n3xt}}. 
\omf{We evaluate \proposal using a wide range of \omf{20} real-world workloads and 7 multi-programmed mixes}.
On systems with 1, 16, 64, and 128 cores, \proposal provides 
\omf{1.2$\times$--2.9$\times$} speedup and \omf{1.2$\times$--1.4$\times$} energy reduction
across all \omf{evaluated} workloads and core counts, with a \nmg{12.3}\% reduction in logic area, compared to the state-of-the-art M3D-based system~\omf{\cite{aly2018n3xt,srimani2023n3xt}, while providing 4.96$\times$ and 7.14$\times$ average speedup compared to the state-of-the-art 3D and 2D systems,
respectively.} 
\omc{As part of our design space exploration, we analyze the impact of \proposal's design decisions for M3D-based systems with various main memory latency values
since 
memory latency can vary depending on the design decisions made to meet certain requirements of the target system. 
This analysis facilitates \omf{making} the \omf{appropriate} design decisions based on the M3D main memory latency.}
\taco{Our analysis demonstrates} that by re-architecting M3D \taco{systems} based on their shifted bottlenecks \nmg{and \taco{unique} opportunities \omf{provided by M3D integration}}, we can \omf{re-}allocate the logic area \omf{to different units in the processor core and caches} more efficiently, \omf{thereby} benefiting a wide range of workloads. 
    
\nmg{This work makes} the following \textbf{key contributions}: 
\squishlist 
    \item We conduct a rigorous design space exploration of key components of the processor core and cache hierarchy to \nmg{understand the implications \omf{of the M3D integration of logic and memory layers on the applications' performance and energy} bottlenecks}. 
    \item We \omf{introduce} \proposal, \omf{an optimized M3D-based system},  based on our insights from our analysis and leveraging M3D \taco{integration's} opportunities, \taco{such as} tightly-integrated logic layers and fast and energy-efficient main memory.
    \item We show that \proposal provides significant performance improvement and energy reduction compared to the state-of-the-art M3D-based system, while achieving a smaller logic area.
\squishend

\section{M3D Technology}
\label{sec:background}

\nmg{We provide a brief overview of the M3D integration technology, state-of-the-art M3D\omf{-based} systems, and their technology feasibility and constraints.}

\begin{figure}[b] %
  \centering
      \vspace{-0.1em}
  \begin{minipage}[b]{0.48\textwidth}
    \centering
    \includegraphics[width=\textwidth]{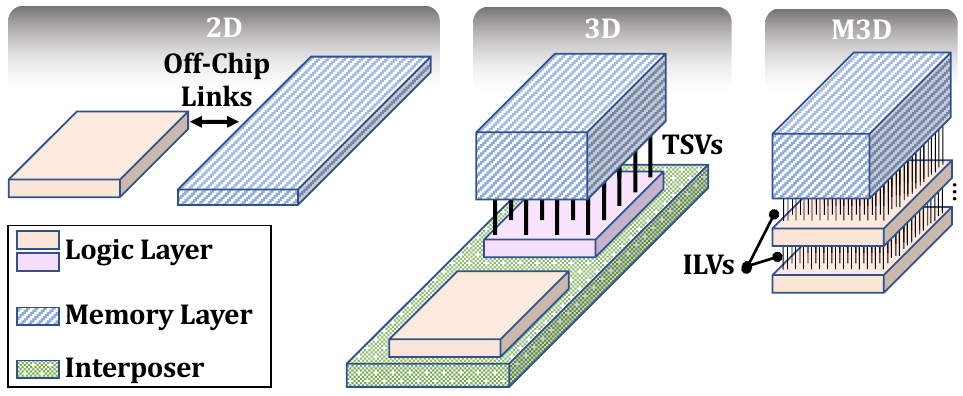}
    \caption{\nmg{Systems with 2D, 3D, and M3D integration.}}
    \label{fig:background-integration}
  \end{minipage}
  \hfill
  \begin{minipage}[b]{0.48\textwidth}
    \centering
    \includegraphics[width=\textwidth]{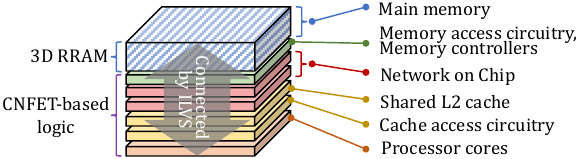}
    \vspace{0.5em}
    \caption{\nmg{Overview of the baseline M3D system~\cite{aly2018n3xt}.}}
    \label{fig:background-n3xt}
  \end{minipage}
  \Description{}
\end{figure}

\head{M3D Integration Technology} There has been significant effort into enabling tighter integration of logic and memory layers in the system to improve performance and energy. 
\fig{\ref{fig:background-integration}} compares three different systems. In 2D systems, computation and memory units are connected via off-chip links with low bandwidth, which is imposed by the limited number of I/O pins in the memory system. In 3D systems~\cite{motoyoshi2009through, mutlu2022modern, ahn2015scalable, boroumand2018google, boroumand2019conda, hsieh2016accelerating,nair2015active,nai2017graphpim,kim2016ultra,niu2022184qps,fujun2020stacked,arnaud2020three,hu2021development,nigussie2021design}, memory layers are connected using Through-Silicon Vias (TSVs) or \taco{hybrid bonding} and can be connected to logic units directly below the memory layers using TSVs/hybrid bonding or on an interposer substrate. Compared to 2D systems, 3D systems provide higher main memory bandwidth, lower latency, and better energy efficiency. In M3D-based systems~\cite{srimani2023n3xt,shulaker2013carbon, walden2021monolithically,shulaker2014monolithic, jagasivamani2020tileable,shulaker2017three, aly2015energy, aly2018n3xt,jagasivamani2019analyzing}, logic and memory layers are \emph{monolithic\omct{ally}-fabricated on the same chip} with a large number of thin \nmg{Inter-Layer Vias (ILVs) with} very small  pitch sizes (e.g., $<$100~nm~\cite{srimani2023n3xt,aly2018n3xt,gopireddy2019designing}). 
Compared to TSV, which has a pitch size of $\sim$2.6--25~$\mu$m~\cite{gopireddy2019designing,kim2021signal}, \taco{and hybrid bonding, which has a pitch size of $\sim$0.3--10~$\mu$m~\cite{kim2016ultra,niu2022184qps,fujun2020stacked,arnaud2020three,hu2021development,nigussie2021design}}, the smaller pitch size of ILVs enables a much denser connectivity.

There are two classes of M3D\omf{-based} systems: \romnum{i}~\emph{only} with logic layers and \romnum{ii} with \emph{both} logic and memory layers. 
The first type provides several optimization opportunities, such as efficient 3D partitioning of logic structures to reduce their latency by shortening the wires~\cite{stow2019efficient,gopireddy2019designing}. The second type enables additional benefits by significantly alleviating main memory bottlenecks~\cite{srimani2023n3xt,aly2018n3xt,zhu2021high,li2021monolithic}. In this work, we consider the second type.
High memory bandwidth is realized by the dense ILVs between logic and memory layers.
\taco{Low memory latency is achieved} due to three reasons.
First, 
\taco{the high-bandwidth ILVs enable efficiently devising a large number of memory channels, each connected to relatively small
memory arrays. The smaller wire lengths in these arrays lower overall access latency, and the larger channel counts lower the contention across concurrent memory accesses. Second, the low pitch size of the ILVs enables efficient M3D layouts of memory arrays (e.g., placing different banks in a memory channel in different layers). This allows further reduction of wire lengths and, subsequently, reduction in the memory array's latency.\footnote{\taco{Note that larger pitch sizes reduce the benefits of fine-grained 3D partitioning of memory arrays~\cite{gopireddy2019designing}.}}}
Third, short and dense ILVs lead to low-latency interconnection between logic and memory on the same chip.

\head{\taco{Baseline} Architecture} We use N3XT~\cite{srimani2023n3xt,aly2018n3xt,aly2015energy, shulaker2017three, shulaker2014monolithic}, a state-of-the-art M3D\omf{-based} system, as our baseline. 
\taco{N3XT leverages emerging technologies such as carbon-nanotube-based transistors in the logic layer~\cite{shulaker2013carbon, shulaker2014monolithic, shulaker2017three} and latency-optimized monolithic 3D RRAM arrays in the memory layers~\cite{kawahara20128, chou2018n40}.}
\taco{The fabrication and thermal feasibility of the N3XT architecture has been demonstrated by real prototypes~\cite{shulaker2017three}.} 
\fig{\ref{fig:background-n3xt}} shows the baseline M3D architecture~\cite{aly2018n3xt}, integrating several logic layers (one for processor cores, one for cache access circuitry, one for shared L2 cache, two for interconnection network to enable high-bandwidth communication between main memory and cores, one for memory access circuitry and memory controllers) and memory layers (64 RRAM layers). Note that the 3D layout of interconnection networks leads to reduced wire lengths and a reduced overall planar footprint~\cite{black2006stacking,kim2007novel,pavlidis20073}. This is achieved by utilizing vertical connections to traverse layers, which are significantly shorter than the planar distances required in 2D layouts.

\taco{In this work, although we use N3XT as our baseline M3D-based system, we primarily focus on characteristics such as main memory bandwidth and latency rather than device-specific features. Because we evaluate M3D-based systems using a range of parameters (e.g., main memory latency), our insights can be broadly applied to enhance performance and energy efficiency in M3D-based systems that employ different logic and memory device technologies.}

\taco{\head{Technological Feasibility and Constraints} Prior work shows that both logic and memory layers in N3XT can be fabricated at \emph{low temperatures} ($<$300$^{\circ}$C~\cite{aly2018n3xt,srimani2023n3xt}). This paves the way to fabricating layers monolithically on top of each other in the same chip without the top layers melting the lower layers during fabrication. Some recent works even demonstrate the feasibility of  M3D fabrication using commercial silicon facilities~\cite{bishop2020fabrication,srimani2023n3xt}. Such works facilitate the wider adoption of M3D integration technology by alleviating the need to devise completely new fabrication lines for M3D. 
Despite all the benefits, there are two key challenges and constraints for M3D fabrication. First, since the M3D fabrication process is sequential (i.e., device layers need to be monolithically and sequentially integrated on top of each other, as opposed to 3D-stacked devices where layers can be fabricated in parallel), M3D technology suffers from concerns typically regarding increased cost and reduced yield. Second, RRAM, the memory technology used in the baseline M3D-based system considered in this work, has limited write endurance. N3XT addresses this issue using online data relocation wear-leveling, extending the device lifetime to at least 10 years~\cite{aly2018n3xt}. We incorporate the same solution in our evaluations (\sect{\ref{sec:methodology}}). We consider the investigation of more efficient solutions (e.g.,~\cite{singh2024morse,bartolo2023mc}) for addressing device challenges as a critical and orthogonal direction.}

\section{Methodology}
\label{sec:methodology}

\subsection{Workloads}

Table~\ref{table:wls} summarizes the wide range of \rev{multi-threaded} workloads \omct{(along with their inputs and settings)} used in our analysis, selected from DAMOV\omct{~\cite{oliveira2021damov,damovgit}}, a benchmark suite for main memory data movement studies. DAMOV aggregates workloads from diverse benchmark suites, including workloads that (1) represent different types of data movement bottlenecks, and (2) come from a wide range of application domains. This is particularly well-suited for our analysis because our goal in this work is to architect the system with monolithically integrated logic and memory layers, where main memory bottlenecks are significantly alleviated.

\begin{table*}[t]
\Copy{R2/4-table}{
\centering
\caption{List of workloads and their input sizes.}
\vspace{-0.1em}
\resizebox{0.96\columnwidth}{!}{
\begin{tabular}{p{3em}cccccccccc}
\toprule
\textbf{Class} & \textbf{Workload} & \textbf{Suite} & \textbf{Domain} & 
\textbf{{Input}} &
\textbf{Size (MB)} & \textbf{BE(\%)} & \textbf{Mem (\%)} & \textbf{BW (\%)} & \textbf{ILP} & \textbf{LFMR}\\ \midrule
\multirow{8}{3.5em}{\omct{{Main} {Memory}} {Bandwidth}-bound}      
& YOLO & Darknet~\cite{redmon_darknet2013} & {Machine Learning} %
& {ref} & 204 & 94.17 & 62.01 & 56.60 & \gf{2.25} & 0.99\\
              & BFS  & Ligra~\cite{ligra} & {Graph Processing} %
              & {rMat} & 2017 & 44.94 & 59.14 & 40.56  & 1.71 & 0.98\\
              & BC   & Ligra~\cite{ligra}& {Graph Processing} %
              & {rMat} & 2017 & 75.72 & 65.16 &  56.84 & 2.09 & 0.99\\
              & KCore & Ligra~\cite{ligra}& {Graph Processing} %
              & {rMat} & 2017 & 94.54 & 44.88 & 78.68 & 1.62 & 0.99\\
              & MIS   & Ligra~\cite{ligra}& {Graph Processing}  %
              & {rMat} & 2017 & 86.70 & 71.72 & 90.88 & 2.02 & 0.99\\
              & PageRank  & Ligra~\cite{ligra}& {Graph Processing} %
              & {rMat} & 2017 & 86.70 & 71.72 &  90.88 & 2.02  & 0.99\\
              & Radii  & Ligra~\cite{ligra}&  {Graph Processing}  %
              & {rMat} & 2017 & 54.12 & 43.10 & 66.34 &  1.78 & 0.99\\ \hline
\multirow{7}{3.5em}{\omct{{Main} {Memory}} Latency-bound} 
& StreamCluster  & Rodinia~\cite{rodinia} & {Data Mining} %
& {ref} & 67 & 63.84& 43.22&  17.38 & 1.74  & 0.99\\
              & ResNet & Darknet~\cite{redmon_darknet2013}& {Machine Learning} %
              & {ref} & 230 & 62.66 & 55.00 & 26.74 & 2.25  & 0.99\\
              & Oceanncp & Splash-2~\cite{splash2} & {HPC} %
              & {simlarge} & 17 & 92.98 & 47.02 & 22.12 & 6.63  & \omct{0.99}\\
              & Components  & Ligra~\cite{ligra} & {Graph Processing}  %
              & {rMat} & 2017 & 50.94 & 42.12  &  6.62  &  1.38 &  0.99\\
              & Triangle   & Ligra~\cite{ligra} & {Graph Processing} %
              & {rMat} & 2017 & 62.08 & 51.10 &  18.74  &  1.41 &   0.99\\
              & Myocyte    & Rodinia~\cite{rodinia} & {Simulation} %
              & {1000000} & 364 & 93.44 & 89.26 & 29.92 & 1.88  & 0.99\\ \hline
\multirow{8}{2em}{{Compute-bound}} & 3mm   & PolyBench~\cite{pouchet2012polybench}& {Linear Algebra} %
& {simlarge} & 128 & 60.3\omct{0} &  13.8\omct{0} &  34.68 & 2.75  & 0.61\\
              & 2mm   & PolyBench~\cite{pouchet2012polybench}& {Linear Algebra} %
              & {simlarge} & 128 & 62.50 & 13.8\omct{0} &  35.29 &  2.55 & 0.60\\
              & atax  & PolyBench~\cite{pouchet2012polybench}& {Linear Algebra} %
              & {simlarge} & 512 & 25.50 & 1.60&   14.90 & 2.37  & 0.51\\
              & gemm  & PolyBench~\cite{pouchet2012polybench}& {Linear Algebra} %
              & {simlarge} & 96 & 63.4\omct{0} & 13.8\omct{0} & 23.11  & 2.55  & 0.58\\
              & ferret  & PARSEC~\cite{parsec} & {Similarity Search} %
              & {ref} & 47 & 29.22 & 4.5\omct{0} & 0.5\omct{0} & 2.64     & 0.61\\
              & Needleman-Wunsch & Rodinia~\cite{rodinia}& {Bioinformatics} %
              & {ref} & 4295 & 79.96& 39.66& 65.46 & 2.35  & 0.52\\
\bottomrule
\end{tabular}}
}
\vspace{-0.4em}
\label{table:wls}
\end{table*}

To gain an understanding of our workload's data movement bottlenecks, we perform a \emph{top-down analysis}~\cite{yasin2014top}, a widely-adopted approach that hierarchically characterizes the workload bottlenecks~\cite{yasin2014top, kanev_isca2015, sirin_damon2017, appuswamy_ipdpsw2018}. 
We use the Intel VTune Profiler~\cite{vtune} to perform the top-down analysis. \taco{For each workload, Table~\ref{table:wls} shows the backend-bound (\texttt{BE}),
DRAM-bound (\texttt{Mem}), and DRAM bandwidth-bound (\texttt{BW}) values obtained from VTune, while running on Intel Xeon E3-1240 processor with 4 cores.
\texttt{BE} accounts for the percentage of execution time during which the processor cannot issue new instructions due to stalls at the pipeline backend, including stalls due to memory requests and core execution. The stall due to memory requests further divides into stalls due to caches or DRAM (\texttt{Mem}). \texttt{Mem} can be further split into DRAM latency- and bandwidth-bound (\texttt{BW}).}  
Table~\ref{table:wls} also shows 
Last-to-First Miss Ratio (LFMR)\omct{~\cite{oliveira2021damov}}, a metric proposed for \omct{evaluating} the effectiveness of caches~\cite{oliveira2021damov}, obtained by dividing the number of L2 cache misses \nmg{(last-level cache in our M3D baseline)} by the number of L1 cache misses. 
A high LFMR suggests that the L2 cache does not provide significant benefits in reducing the miss rate.  We consider LFMR above 90\% as high, as most requests are not serviced by the L2 cache.

Based on this characterization, we classify the workloads into \emph{main memory bandwidth-bound}, \emph{main memory latency-bound}, and \emph{compute-bound}. \taco{We categorize the workloads with large \texttt{BE} ($>$40\%), \texttt{Mem} ($>$40\%), and \texttt{BW} ($>$50\%) as DRAM bandwidth-bound since we observe large performance improvements when improving the main memory bandwidth for these workloads. We categorize workloads with high \texttt{BE} and \texttt{Mem} and low \texttt{BW} as DRAM latency-bound since we observe large performance improvements when improving the main memory latency for these workloads. We categorize workloads with high \texttt{BE} and low \texttt{Mem} as compute-bound.
When a workload nearly misses the border between main memory-/compute-bound, we use LFMR to classify the workload as memory-bound (LFMR$>$90\%) or compute-bound.}

\nmg{For our end-to-end evaluations, we also evaluate a wide range of multi-programmed workload mixes, with a focus on combining workloads with various bottlenecks.}
\taco{We summarize the multiprogram workload mixes in Table~\ref{table:multi}}.
\begin{table}[h]

\caption{Multi-programmed workload mixes.\vspace{-0.3em}}
\label{table:multi}
\resizebox{0.38\columnwidth}{!}{
\begin{tabular}{lll}
\toprule
Class                                                                                    & \omct{ID} & Workloads                \\ \hline
\multirow{4}{*}{\begin{tabular}[c]{@{}l@{}}Memory-bound + \\ Compute-bound\end{tabular}} & 1   & Components + 3mm         \\
                                                                                         & 3   & MIS + atax               \\
                                                                                         & 6   & PageRank + 2mm           \\
                                                                                         & 7   & Triangle + gemm          \\ \hline
\begin{tabular}[c]{@{}l@{}}Memory-bound +\\ Memory-bound\end{tabular}                    & 5   & Oceanncp + StreamCluster \\ \hline
\multirow{2}{*}{\begin{tabular}[c]{@{}l@{}}Compute-bound +\\ Compute-bound\end{tabular}} & 2   & gemm + atax              \\
                                                                                         & 4   & NW + 3mm                 \\ \bottomrule
\end{tabular}
}
\end{table}

\subsection{Experimental Setup}

\newcolumntype{?}{!{\vrule width 1pt}}
\newcolumntype{;}{!{\vrule width 0.5pt}}
\begin{table*}[b]
\centering
\caption{2D, \new{3D}, and M3D system configurations.}
\label{table-parameters2}
\resizebox{0.9\columnwidth}{!}{
\begin{tabular}{l?lll}
\toprule
\textbf{System} & M3D & \new{3D} & 2D \\ 
\hline
\multicolumn{1}{l?}{\begin{tabular}[c]{@{}l@{}}\textbf{Main}\\ \textbf{Memory}\end{tabular}} & 
\begin{tabular}[c]{@{}l@{}} 64 GB on-chip 3D RRAM~\cite{aly2018n3xt};
\\ \new{Up to} 16 TB/s~\cite{aly2018n3xt}; \\ 5/13ns \nmg{R/W} \taco{\emph{(up to 5/60ns R/W)}};\\  0.8/1.1 pJ/bit \nmg{R/W}~\cite{aly2018n3xt}\end{tabular} &
\begin{tabular}[c]{@{}l@{}} 64 GB on-chip 3D-stacked DRAM using \\ 5um TSVs; HBM2 interface, 1.5 TB/s~\cite{hbm3skhynix,hbmA100}; \\  51/55ns \nmg{R/W};\\ 9~pJ/bit~\cite{aly2018n3xt} \end{tabular} &
\begin{tabular}[c]{@{}l@{}} 64 GB off-chip DRAM;\\ DDR4 interface,  102 GB/s~\cite{ghose.sigmetrics20};\\ 65/60ns \nmg{R/W}~\cite{aly2018n3xt};\\ 20pJ/bit \nmg{R/W}~\cite{mailthody2019deepstore} \end{tabular}  \\ 
\hline
\multicolumn{1}{l?}{\textbf{L3}} & None & None & \begin{tabular}[c]{@{}l@{}}Shared 8~MB, 16-way, 27-cycle;
\\
945/1904 pJ per hit/miss \cite{tsai2018adaptive}\\\end{tabular} \\ 
\hline
\multicolumn{1}{l?}{\textbf{L2}}  & 
\begin{tabular}[c]{@{}l@{}}Shared 256~KB per core \emph{(0/1/8/64~MB)}\\ 8-way, 12-cycle \emph{(6-cycle)};\\ 46/93 pJ per hit/miss \end{tabular} &       \begin{tabular}[c]{@{}l@{}}Shared 256~KB per core,\\ 8-way, 12-cycle;\\ 46/93 pJ per hit/miss \end{tabular} &  \begin{tabular}[c]{@{}l@{}}Private 256~KB per core,\\ 8-way, 12-cycle;\\ 46/93 pJ per hit/miss \end{tabular} \\
\hline
\multicolumn{1}{l?}{\textbf{L1}}  & \multicolumn{3}{l}{\begin{tabular}[c]{@{}l@{}} 32~KB, 8-way, 4-cycle \emph{(2-cycle)}; 15/33 pJ per hit/miss~\cite{tsai2018adaptive} \end{tabular}} \\ 
\hline
\multicolumn{1}{l?}{\begin{tabular}[c]{@{}l@{}}\textbf{Processor}\\ \textbf{Core}\end{tabular}}  & \multicolumn{3}{l}{\begin{tabular}[c]{@{}l@{}} 1/16/64/128 out-of-order cores @ 4~GHz; 4-wide pipeline \emph{(8-wide)}; 128-entry reorder buffer \emph{(256- and 512-entry)};\\
Branch predictor: two-level GAs~\cite{yeh1991two} \emph{(TAGE-SC-L~\cite{seznec2016tage} and ideal predictor with no mispredictions)};\\ 0.48 nJ/inst for M3D; 1.5 nJ/inst for 2D and 3D \end{tabular}} \\
\bottomrule
\end{tabular}
}
\end{table*}

\nmg{\head{Modeled Systems}} We use a cycle-accurate architectural simulator (ZSim~\cite{sanchez2013zsim}) to flexibly explore various designs and faithfully model 
the following systems:
\squishlist
    \item \textbf{2D:} a system based on the 2D integration of computation and memory, where the cores and main memory (DRAM) communicate through an off-chip bus.
    
    \item \textbf{3D:} a system that uses TSVs to integrate memory layers, which connect to compute units via interposers (e.g., \cite{hbm3skhynix}). We consider six HBM stacks connected to compute units similar to a state-of-the-art 3D system~\cite{NVIDIA-A100}.
    
    \item \textbf{M3D:} a system based on N3XT~\cite{srimani2023n3xt,aly2018n3xt,aly2015energy, shulaker2017three, shulaker2014monolithic}, where memory and logic are monolithically fabricated on each other. \omf{We}  experimentally calibrate the M3D-based system parameters \omf{based on} the fabricated device models, circuit analysis, interconnect model, and physical design tools presented in ~\cite{aly2018n3xt}.
    \taco{The latency of on-chip main memory in the M3D-based system can depend on various factors (e.g., memory capacity, size of each memory array, and length of local and global wires in the memory array). Therefore, to gain comprehensive insights into M3D-based system design implications, we also evaluate M3D-based systems with varying memory latency values in our design space exploration (\sect{\ref{sec:architecting}}) and end-to-end evaluations (\sect{\ref{sec:revamp-eval}}). To this end, we rigorously sweep the memory latency values from half the value reported for the baseline M3D configuration  up to a latency value similar to the baseline 2D system.}

\squishend

 We summarize all system parameters in Table~\ref{table-parameters2}.
 \footnote{The main memory latency values in this table refer to average main memory access latency.} 
 We also show other configurations analyzed in \sect{\ref{sec:architecting}} in parentheses.

\nmg{\head{Cache and Core}} We initially consider the same core and L1 caches for all configurations. 2D has a deeper \nmg{and larger} cache hierarchy compared to 3D and M3D since it has the lowest available memory bandwidth. We consider a cache hierarchy for M3D  similar to N3XT~\cite{aly2018n3xt} (private L1 and shared L2). We consider the same cache hierarchy for 3D. 
In \sect{\ref{sec:architecting}}, we study the effects of different cache and core configurations in M3D.

\nmg{\head{Main Memory}} 
In our analysis, all three systems have large enough main memory capacity to \taco{contain the working set of all applications in Table~\ref{table:wls}}.\footnote{\appmove{We also discuss the case where data does not fit in one stack in \sect{~\ref{sec:discussion}.}}}
Although the main memory in the M3D baseline is RRAM-based and different from DRAM used in 2D and 3D, in our performance analysis, we primarily distinguish memory systems by their bandwidth and latency and not their other device features. 
\taco{As explained in \sect{\ref{sec:background}}, in our evaluations, we employ the same technique used in the M3D baseline for alleviating the reliability and endurance issues of the RRAM-based memory~\cite{aly2018n3xt}.}

\nmg{
\head{Network on Chip}
The M3D baseline~\cite{aly2018n3xt} uses the meshes of trees (MoT)~\cite{balkan2009mesh} scheme to provide a high-bandwidth connection between the cores and the memory controllers. This network supports \omct{equal-hop-count paths from every core to every memory channel.}
}

\section{\nmicro{Motivational Analysis}}
\label{sec:bottlenecks}

We conduct an experimental analysis of performance and energy bottlenecks of a variety of workloads on a state-of-the-art M3D-based system.

\subsection{Performance Bottlenecks}
\label{sec:bottlenecks-performance}

\begin{figure}[h]
\Copy{R3/1a-fig}{
\centering
\includegraphics[width=0.4\linewidth]{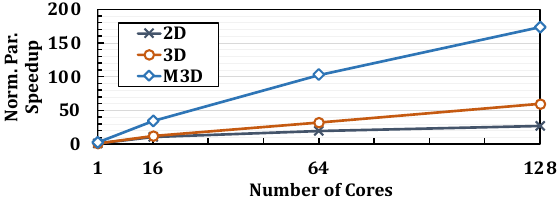}
\vspace{-0.3em}
\caption{Speedups of the M3D system over the 3D and 2D systems \omct{across all our evaluated workloads}.}
}
\label{fig:motivation-new}
\Description{}
\end{figure}

We evaluate the performance benefits of the M3D-based system over the 2D and 3D systems across \emph{all our \omct{20 evaluated real-world workloads and 7 multi-programmed mixes}}. \fig{\ref{fig:motivation-new}} shows the performance of 2D, 3D, and M3D baselines averaged across all our workloads. 
\omct{Performance refers to the normalized parallel speedup, where all values are normalized against that of the 2D baseline with one core.}
We observe that the M3D system provides significant benefits over 2D and 3D baselines, delivering 4.0$\times$ and 2.8$\times$ average speedups over 2D and 3D systems, respectively.

\begin{figure}[b] %
  \centering
  
  \begin{minipage}[b]{0.48\textwidth}
    \centering
    \includegraphics[width=\textwidth]{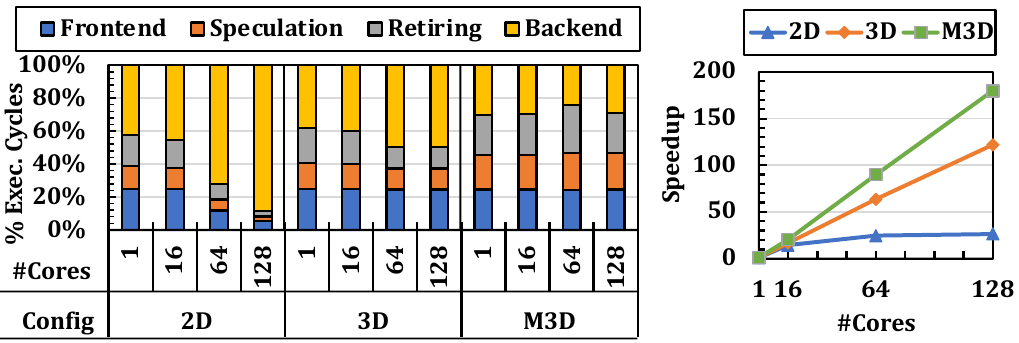}
    \caption{\nmg{Bottlenecks of a \omct{main memory} latency-bound workload \omct{(\textsf{Tri})}.}}
    \label{fig:bottlenecks-tri}
  \end{minipage}
  \hfill
  \begin{minipage}[b]{0.48\textwidth}
    \centering
    \includegraphics[width=\textwidth]{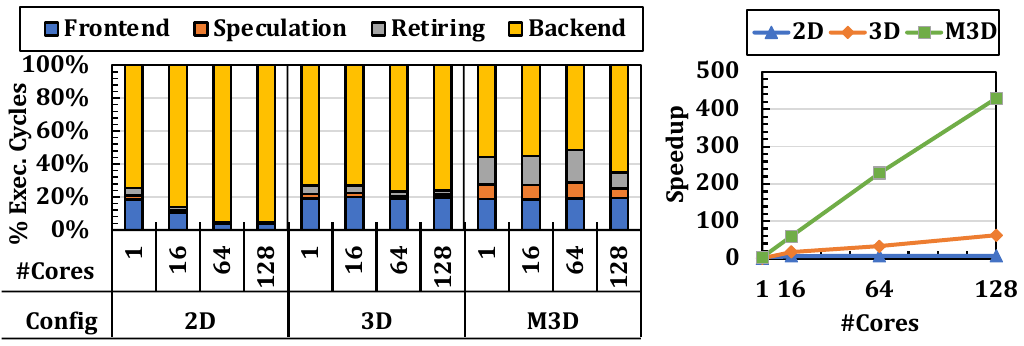}
    \caption{\nmg{Bottlenecks of a \omct{main memory} bandwidth-bound workload \omct{(\textsf{BFS})}.}}
    \label{fig:bottlenecks-bfs}
  \end{minipage}
\Description{}
\end{figure}

To show the impact of M3D on the performance bottlenecks \nmg{of the conventionally main memory-bound workloads}, we compare the performance bottlenecks of \emph{memory latency-/bandwidth-bound} workloads.  To this end, we extend ZSim \cite{sanchez2013zsim} by incorporating the top-down bottleneck analysis~\cite{yasin2014top}. 
\taco{We validate our implementation of the top-down analysis in ZSim by comparing its output against Intel VTune~\cite{vtune} running on Intel Xeon E3-1240. We show that the ZSim-based top-down analysis identifies very similar \emph{trends} in bottlenecks to Vtune~\cite{vtune}, with a high Pearson correlation~\cite {benesty2009pearson} of 93.94\% ($P<0.001$) across our workloads.} 
\fig{\ref{fig:bottlenecks-tri}} 
\revtaco{and} \fig{\ref{fig:bottlenecks-bfs}}
show top-down bottleneck breakdown of representative latency-bound (\texttt{Tri}) and a bandwidth-bound (\texttt{BFS}) workloads.
In M3D, we observe a large reduction \grm{i}n the ratio of execution time spent on the backend, compared to 2D and 3D,
and a \omf{substantial} \emph{bottleneck shift} in the system from the backend to the frontend and speculation.

\taco{To gain a deeper understanding of the backend bottlenecks (which are influenced by the main memory, cache hierarchy, and the processor's functional units), we study an \emph{idealized} M3D-based system with a one-cycle main memory latency and no main memory bandwidth bottleneck. We observe that even such a configuration only leads to a 23\% speedup, lowering the backend bottleneck only by 5\%, which means that the main memory is not a key contributor to the M3D-based system's backend bottleneck.}
We conclude that the advancements in M3D memory performance result in substantially \emph{shifting}  the application bottlenecks from main memory to other parts of the system.

\noindent\ch{lavendergray}{\textbf{Takeaway 1.}}
\emph{The advancements in M3D memory performance \omct{can} substantially shift the applications' performance bottlenecks from main memory to other parts of the system.}

\subsection{\nmg{Energy Bottlenecks}}
\label{sec:bottlenecks-power}

We analyze the energy bottlenecks in different system\omct{s} and observe that,
compared to the 2D (3D) systems, the M3D-based system consumes on average 4.32$\times$ (4.76$\times$) and 4.13$\times$ (3.32$\times$) lower energy on compute-bound and memory-bound workloads, respectively. 
Even though for the memory-bound workloads, the main memory consumes a significant percentage of the overall energy in 2D and 3D, the main memory in the M3D baseline contributes to only 12\% of total energy, turning the processor core into the most prominent contributor to the total energy consumption in M3D. 

\taco{The energy benefits of the M3D baseline are due to two key factors. The first factor relates to device-level advances in logic and memory layers, which, in fact, are orthogonal to the M3D integration. For example, the CNFET-based memory access circuits consume less power compared to MOSFETs~\cite{shulaker2013carbon}, and the deployed RRAM-based memory layers are more energy-efficient than DRAM~\cite{shulaker2017three, aly2015energy, aly2018n3xt}. The second factor relates to the M3D integration. As discussed in \sect{\ref{sec:background}}, M3D integration efficiently supports many memory channels, each connected to smaller memory arrays.  Due to the shorter wire lengths within the memory arrays and between memory and logic layers on the same chip, the overall energy of memory accesses gets lowered with M3D integration.}

\noindent\ch{lavendergray}{\textbf{Takeaway 2.}}
\emph{In the M3D \omct{system}, the processor core turns into \omct{a} critical energy bottleneck.}

\subsection{Our Goal}
Given the opportunities provided by M3D to drastically shift the performance and energy bottlenecks, it is important to understand the implications of this new technology on \omf{processor core and cache hierarchy} designs that have been conventionally specialized to tackle main memory bottlenecks. \textbf{Our goal}  in this work is to design the core and cache\revtaco{s}, given the new trade-offs of M3D. 
While we focus on architecting M3D systems in this work, we argue that if \taco{other integration technologies (e.g., TSV or hybrid bonding) in the future achieve the same properties (e.g., pitch sizes, \omct{latencies, and bandwidths}) as current M3D technologies}, our architectural insights can also be applied to systems \taco{with such integration technologies}. 

\section{Architecting the Processor Components}
\label{sec:architecting}

To design general-purpose processors based on the new opportunities and shifted bottlenecks of M3D-based systems, we analyze the design trade-offs of the cache hierarchy (\sect{\ref{sec:arch-caches}}) and cores (\sect{\ref{sec:arch-cores}}) in M3D. Table~\ref{table:structure} summarizes \romnum{i}~the key structures analyzed in this work, in
different stages of \omf{an out-of-order superscalar} processor core pipeline,
\romnum{ii}~the key analyzed aspect of each structure, and \romnum{iii}~a pointer to the relevant sections.

\begin{table}[h]
\caption{Analyzed pipeline elements.\vspace{-0.5em}}
\centering
{
\resizebox{0.5\columnwidth}{!}{%
\footnotesize
\begin{tabular}{llll}
\toprule
\textbf{Stage} & \textbf{Key Structure} & \textbf{Analyzed Aspect} & \textbf{\omct{Section}} \\ \midrule
Fetch & \begin{tabular}[c]{@{}l@{}}Instruction cache \\ Branch predictor\end{tabular} & \begin{tabular}[c]{@{}l@{}}Part of frontend bottleneck\\ Speculation bottleneck\\ Part of pipeline width\end{tabular} & \begin{tabular}[c]{@{}l@{}} \sect{\ref{sec:arch-spec}} \\ \sect{\ref{sec:arch-spec}} \\ \sect{\ref{sec:core-wide}} \end{tabular} \\ \midrule
Decode                & \begin{tabular}[c]{@{}l@{}} Pre-decoder\\ Decoder\end{tabular}           & \begin{tabular}[c]{@{}l@{}} Part of frontend bottleneck\\ Part of pipeline width\end{tabular}                                  & \begin{tabular}[c]{@{}l@{}}  \sect{\ref{sec:arch-spec}} \\ \sect{\ref{sec:core-wide}} \end{tabular} \\ \midrule
Issue & Instruction window & Part of pipeline width& \sect{\ref{sec:core-wide}} \\ 
 & Reorder buffer &  Reorder buffer size & \sect{\ref{sec:queue-size}} \\ \midrule
Execution             & Functional units                                                                        & \begin{tabular}[c]{@{}l@{}}Part of pipeline width\\ Latency of micro-operations\end{tabular}                         & \begin{tabular}[c]{@{}l@{}}\sect{\ref{sec:core-wide}}\\ \sect{\ref{sec:uop}} \end{tabular} \\ \midrule
Memory                & \begin{tabular}[c]{@{}l@{}} L1/L2 data caches\\ Load/store queues\\ Threads communication \end{tabular} & \begin{tabular}[c]{@{}l@{}} Cache size, latency, depth\\ Load/store queue sizes\\ Synchronization\end{tabular}                                                                          & \begin{tabular}[c]{@{}l@{}}\sect{\ref{sec:arch-caches}}\\ \sect{\ref{sec:queue-size}}\\ \sect{\ref{sec:sync-opt}}\end{tabular}     \\ \midrule
Writeback            & Register file (RF)                                                                           & Part of pipeline width                                                                                                                                                         & \sect{\ref{sec:core-wide}}                                                 \\ \bottomrule
\end{tabular}}}
\label{table:structure}
\vspace{-5pt}
\end{table}

\subsection{Cache Hierarchy}
\label{sec:arch-caches}

The design space of the cache hierarchy in M3D-based systems is affected by two conflicting factors. \taco{On one hand, M3D integration enables \omf{reducing the latency of} caches through efficient 3D layouts (e.g.,~\cite{srinivasa2018monolithic,srinivasa1,srinivasa2,kong2017architecting}) that reduce wire lengths. On the other hand, as discussed in \sect{\ref{sec:background}}, M3D integration enables improving main memory performance (bandwidth and latency), which reduces the impact of caches on the application performance. We evaluate the impact of these conflicting factors on the performance of a wide range of workloads running on the M3D-based system.}

\subsubsection{Depth of the Cache Hierarchy.} 
\label{sec:cache-depth}

\begin{figure}[h]
\Copy{R3/1b-fig}{
\centering
\includegraphics[width=0.58\linewidth]{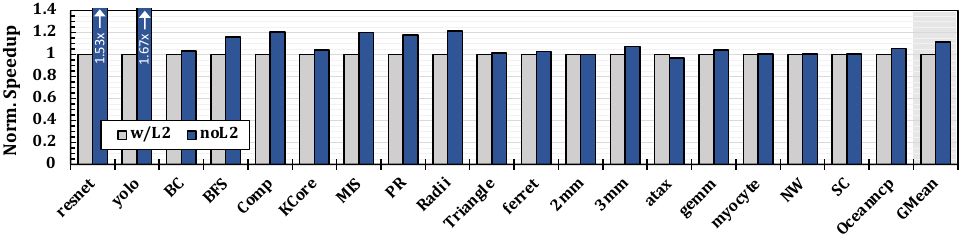}
\vspace{-0.5em}
\caption{Performance of the M3D-based systems with and without the L2 cache \omct{across our evaluated workloads}.}
}
\label{fig:r3-bypass}
\Description{}
\end{figure}

We compare \texttt{noL2}, an M3D-based system with no L2 cache\footnote{\raggedright\nmg{Coherenc\omct{e} is still maintained using a shared directory.}}, and  \texttt{w/L2}, an M3D-based system with a shared L2 cache. \fig{\ref{fig:r3-bypass}} shows the performance of \texttt{noL2} normalized to \texttt{w/L2} across all our workloads on an M3D system with 64 cores, showing on average 11.5\% speedup of \texttt{noL2} over \texttt{w/L2}.

\begin{figure}[b] %
  \centering
      \vspace{0.5em}
  
  \begin{minipage}[b]{0.48\textwidth}
    \centering
    \includegraphics[width=\textwidth]{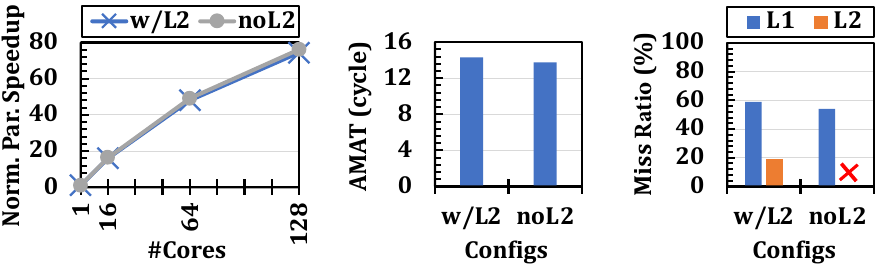}
    \caption{Effect of no L2 cache on a low-LFMR workload \omct{(\textsf{atax})}.}
    \label{fig:depth-2mm}
  \end{minipage}
  \hfill
  \begin{minipage}[b]{0.48\textwidth}
    \centering
    \includegraphics[width=\textwidth]{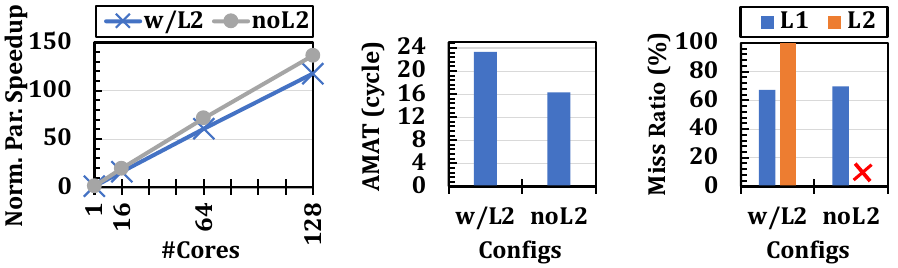}
    \caption{Effect of no L2 cache on a high LFMR workload \omct{(\textsf{MIS})}.}
    \label{fig:depth-MIS}
  \end{minipage}
\Description{}
\vspace{-5pt}
\end{figure}

To provide a more detailed analysis of the impact of removing the L2 cache,
\fig{\ref{fig:depth-2mm}} shows the performance of a representative workload with low LFMR (\texttt{atax}) alongside the Average Memory Access Time (AMAT) and cache miss rates for a 64-core system. We observe that, despite having a low L2 cache miss rate of 19\% in \texttt{w/L2}  and a faster L2  compared to the M3D main memory, removing the L2 cache does not hurt performance.
Second, we observe that the AMAT in \texttt{noL2} is on par (even 4\% lower) with the AMAT in \texttt{w/L2}. While removing L2 does not significantly hurt AMAT for workloads with high L2 hit rates in M3D, removing it in the 3D baseline can lead to an up to 17$\times$ larger AMAT due to its slower main memory. 
\fig{\ref{fig:depth-MIS}} shows the performance, AMAT, and cache miss rates of a representative workload with high LFMR (\texttt{MIS}) for \nmg{the same configurations}. Based on this figure, we make three observations.
First, \texttt{noL2} improves performance by 17.8\% on average across all core counts compared to both configurations with L2. Second, the L2 cache is highly ineffective (99\% misses) in filtering accesses to the main memory. Third, the AMAT in  \texttt{noL2}  is significantly lower than the two other configurations since it does not add extra L2 access latency and contention overhead.

\noindent\ch{lavendergray}{\textbf{Takeaway 3.}}
\emph{In the M3D \omct{system}, removing the shared L2 cache does not degrade performance, and even leads to improved performance for some workloads.}

\begin{figure}[h]
\centering
\includegraphics[width=0.4\linewidth]{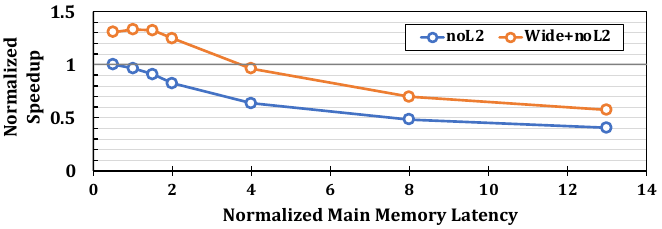}
\caption{Performance impact of removing the L2 cache on M3D-based systems with different memory latency values \omct{for a representative low-LFMR workload (\textsf{atax}). Main memory latency values are normalized to the baseline M3D memory latency of 5 ns}.}
\label{fig:cache-sense-lat}
\Description{}
\vspace{-5pt}
\end{figure}

Due to the importance of analyzing the impact of design decisions on an M3D system with different latency values (as discussed in \sect{\ref{sec:methodology}}), we demonstrate the impact of cache hierarchy for an M3D system with different memory latency values. We keep the memory bandwidth the same (as listed in Table~\ref{table-parameters2}) across configurations. \fig{\ref{fig:cache-sense-lat}} shows the performance of a low-LFMR workload \omct{(\textsf{atax})} on an M3D-based system with 64 processor cores in two different cases: \romnum{i}~a configuration with no L2 (noL2), and \romnum{ii}~a configuration with no L2, and where the area originally dedicated to L2 is used to increase the processor core's pipeline width (Wide+noL2), as explained in \sect{\ref{sec:core-wide}} and \sect{\ref{sec:case-study}}. For each latency value, performance is normalized to the M3D baseline (with L2 cache and baseline pipeline width)
with that latency. We observe that even until 4$\times$ larger latency values, it is beneficial to use the logic area for increasing processing capability (e.g., wider pipeline) instead of larger caches. This is because the large M3D bandwidth enables serving a larger number of in-flight requests. However, for larger latency, having larger caches becomes critical. In \sect{\ref{sec:revamp-eval}}, we analyze the impact of all of \proposal's design decisions in M3D systems with different latency values across all our workloads.

\noindent\ch{lavendergray}{\textbf{Takeaway 4.}}
\emph{Even \omct{with} 4$\times$ larger main memory latency values (compared to the M3D baseline), it is beneficial to use the logic area for increasing processing capability (e.g., a wider pipeline) instead of increasing the cache sizes. This is because the M3D main memory enables efficient serving of a large number of in-flight requests. However, for higher latency values \omct{(i.e., beyond 4$\times$ larger compared to the M3D baseline)}, having larger caches becomes critical.}
\vspace{0.5em}

\head{Microarchitectural implications} We detail the architectural implications of removing the L2 cache in the M3D system by explaining four key aspects. First, removing the shared L2 cache increases main memory accesses for some workloads, but the wear-leveling provisions in the M3D baseline remain sufficient for maintaining lifetime guarantees even in this case. This is because the M3D baseline is already designed to meet its reported lifetime guarantees under the worst-case scenarios: workloads where caches are ineffective at filtering main memory accesses. This is common in memory-bound workloads, as also observed in our evaluations (characterized by high LFMR values in Table~\ref{table:wls}). While compute-bound workloads incur more memory accesses when the L2 cache is removed, their \emph{total access count} remains well below that of memory-bound workloads. To confirm this, we analyzed the maximum number of writes per cycle across all workloads, with and without the L2 cache. We found that the maximum writes/cycle (2.42 for the 64-core configuration) is \emph{unchanged} in both cases. This is because the worst-case write rate originates from the highly memory-bound workload BC, which has low access locality and an L2 miss rate of 99\%. In contrast, compute-bound workloads, which benefit from higher data locality, show significantly lower writes/cycle even without L2 (up to 8x lower than BC). Since removing L2 does not worsen the worst-case write behavior, the baseline wear-leveling technique~\cite{aly2018n3xt} remains sufficient to achieve the same endurance as the M3D baseline~\cite{srimani2023n3xt,aly2018n3xt,aly2015energy, shulaker2017three, shulaker2014monolithic}.

Second, removing the L2 cache in the M3D baseline does not introduce notable energy penalties, and in fact, our overall design (as evaluated in \sect{\ref{sec:revamp-eval}}) improves energy efficiency.
For compute-bound workloads, removing the L2 cache increases the number of memory accesses and the activity of the memory controller. However, in these workloads, given M3D's energy-efficient main memory, most of the energy consumption is due to the processing units (as shown in \sect{\ref{sec:bottlenecks-power}}). Furthermore, the M3D baseline's memory controller consumes very little energy ($\sim$0.01 pJ/bit per controller)~\cite{aly2018n3xt}. For memory-bound workloads, removing L2 caches leads to only an intangible impact since the L2 cache is highly ineffective at filtering the memory accesses. Overall, as demonstrated by the RvM3D-P configuration in \fig{\ref{fig:perf-energy}} (considering only our proposed performance \omct{but not energy} optimizations), removing the L2 caches leads to only a modest 2\% increase in average energy consumption.  Finally, when considering both our proposed performance and energy optimizations, the overall design (RvM3D in \fig{\ref{fig:perf-energy}}) results in even lower (by 1.2$\times$–1.4×$\times$) energy consumption than the M3D baseline.

Third, removing the L2 cache naturally disables L2 prefetching; however, this has minimal impact on performance. This is because, as observed in this section, even for workloads with high L2 hit rates, the absence of the L2 cache leads to comparable performance \omct{to the M3D system with the L2 cache} due to M3D's highly efficient main memory. L1 prefetching remains possible in this configuration, which can be used for workloads that benefit from it.

\begin{figure}[h]
\Copy{R1/1-4fig}{
\centering
\includegraphics[width=0.45\linewidth]{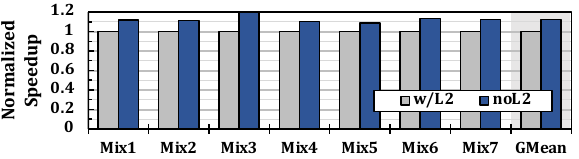}
\caption{Effect of removing the L2 cache on context switches \omct{across our  multi-programmed workloads}.}
}
\label{fig:context-switch}
\Description{}
\end{figure}

Fourth, removing the L2 cache does not impact the context switching process itself, as the contents of the caches are not \omct{saved} during context switches. While the presence of the L2 cache may preserve some program data in the cache across context switches, as shown in this section, even for workloads with high cache hit rates, the L2 cache provides minimal performance benefit (and on average, across all workloads, removing L2 leads to speedups). 
We demonstrate the impact of removing the L2 cache on context-switching by comparing 16-core M3D systems with and without the L2 cache. \fig{\ref{fig:context-switch}} shows the performance of the multi-program workload mixes in Table~\ref{table:multi}, with each workload in the mix using 16 threads. \omct{Performance is normalized to the baseline M3D configuration with the L2 cache.} We observe that the configuration without L2 performs on average 12.7\% better than the configuration with L2.

\subsubsection{Impact of Cache Size.}
\label{sec:cache-size}
We analyze the performance implication\omct{s} of increasing the L2 cache size. 
\nmg{M3D facilitates circuit-level and device-level optimizations to increase cache size or reduce cache latency by designing vertical layouts of SRAM arrays to reduce the overall wire lengths.}
Analyzing the effects of such optimizations \nmg{on M3D-based systems with logic and memory layers} enables us to decide whether removing L2 \nmg{in such systems} is ultimately better than optimizing it. We analyze the performance of workloads with low and high LFMR, with varying L2 sizes 
and no L2. 
Even though increasing cache size \js{can increase} latency, we \nmg{optimistically} consider the ideal case where the latency does not increase with the cache size.  
Across \emph{all \omf{evaluated} workloads} with different core counts, increasing the cache size \omct{by four times} leads to only a 3.7\% average speedup.

\noindent\ch{lavendergray}{\textbf{Takeaway 5.}}
\emph{In the M3D \omct{system}, increasing the L2 cache size does not significantly outperform removing it.}

\subsubsection{Impact of Cache Latency.}
\label{sec:cache-lat}

We study the effect of reducing the latency of L1 and L2 caches. We analyze M3D-based system configurations where \romnum{i}~the L1 latency is improved by 2$\times$ (\texttt{L1fast}), and \romnum{ii}~the L2 latency is improved by 2$\times$, and the L2 size is 64MB (\texttt{L2Opt}). 
Overall, we observe 12.5\%/6\% average speedup due to L1/L2 latency reduction across \emph{all \omf{evaluated} workloads}. We conclude that removing the shared L2 cache leads to better or comparable performance compared to larger or faster L2 caches. However, it is still important to reduce L1 latency. \nmg{This is because} L1 cache latency represents a large portion of the AMAT in M3D due to M3D's fast main memory.

\noindent\ch{lavendergray}{\textbf{Takeaway 6.}}
\emph{In the M3D \omct{system}, removing the shared L2 cache leads to performance \omct{better or comparable to that with} faster L2 caches. However, it is still important to reduce L1 cache latency.}

\subsection{Processor Core Design}
\label{sec:arch-cores}

\subsubsection{\mrev{I}mpact of \mrev{P}ipeline \mrev{W}idth.}
\label{sec:core-wide}

\begin{figure}[h] %
  \centering
  
  \begin{minipage}[b]{0.55\textwidth}
  \Copy{R3/1c-fig}{
    \centering
    \includegraphics[width=\textwidth]{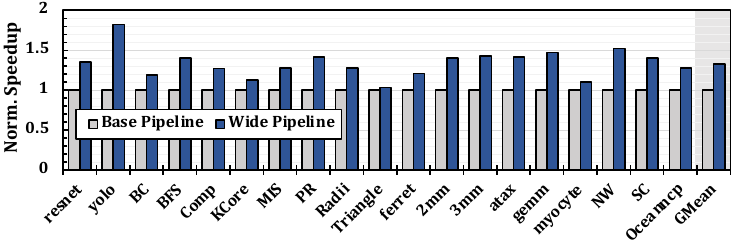}
    \caption{Speedup of an M3D system with a wider pipeline \omct{across our evaluated workloads}.}
    }
    \label{fig:r3-wide}
  \end{minipage}
  \hfill
  \begin{minipage}[b]{0.4\textwidth}
    \centering
    \includegraphics[width=\textwidth]{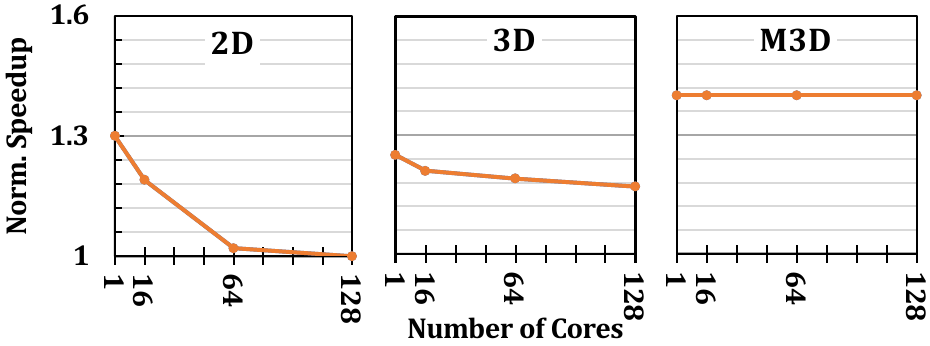}
    \caption{Speedup of the wide pipeline for 2D, 3D, M3D \omct{for a representative main memory bandwidth-bound workload (\textsf{BFS})}.}
    \label{fig:wide-bw-bfs}
  \end{minipage}
\Description{}
\end{figure}

We aim to understand the impact of wider pipelines in the M3D-based system. 
Wider pipelines lead to a higher number of in-flight requests, which can be beneficial when the main memory can serve requests efficiently. 
\fig{\ref{fig:r3-wide}} shows the performance of a 64-core M3D system with doubled pipeline width normalized to the performance of a 64-core baseline M3D system. Across all our workloads, we observe a 32.3\% average speedup.

To differentiate the impact of pipeline width on the performance of 2D, 3D, and M3D-based systems, in \fig{\ref{fig:wide-bw-bfs}}, we show an example of a \omct{main memory} bandwidth-bound workload (BFS) on these
systems when doubling the pipeline width.
Each data point is the speedup of each configuration after widening the pipeline over \emph{the same configuration}. 
We observe that doubling the pipeline width provides a 40\% speedup in the M3D-based system, while providing a lower speedup in the 2D and 3D systems. 
\nmg{This is because, \omct{especially} with large core counts,} main memory \nmg{bandwidth becomes} a major \nmg{performance} bottleneck in the 2D and 3D systems, \omct{thereby leaving these systems unable to efficiently serve higher numbers of requests from the wider pipeline.}

We also analyze the effects of increasing the widths of different pipeline stages \nmg{separately} to understand whether significant benefits are obtained at lower area costs. We observe that doubling the width of \emph{both} the backend and frontend leads to \nmg{at least 1.27$\times$ higher average speedup (across all our evaluated workloads) than doubling the widths individually}. \taco{As discussed in \sect{\ref{sec:cache-depth}}, \fig{\ref{fig:cache-sense-lat}} shows the performance of a wider pipeline on M3D-based systems with high memory bandwidth but varying memory latency values.}

\noindent\ch{lavendergray}{\textbf{Takeaway 7.}}
\emph{Wider pipelines provide large benefits to the M3D \omct{system}.}

\subsubsection{Size of Reordering Structures and Queues.} 
\label{sec:queue-size}
We analyze the performance of the wide pipeline with \romnum{i}~baseline load/store queues (L/SQ), issue queue, and reorder buffer (ROB) depths and \romnum{ii}~two times deeper queues. Across all evaluated workloads, larger queue sizes have a lower performance impact on the M3D-based system compared to the 3D system (on average, 12\% in M3D versus 25\% in 3D) due to lower instruction wait times in the M3D-based system. Despite the relatively lower impact of queue sizes in the M3D-based system, some workloads still take significant advantage of the larger queues (e.g., up to 20\%). 

We further increase the ROB size to 512 entries and continue to observe relatively low average speedup, but with particularly larger benefits for a subset of applications. Relative to a configuration with a 256-entry ROB, this results in speedups of up to 17\%, with an average improvement of 3.5\% across all our evaluated workloads.

\noindent\ch{lavendergray}{\textbf{Takeaway 8.}}
\emph{Although \omct{larger load/store queues, issue queue, and reorder buffer} provide a smaller average performance benefit in the M3D-based system than in 2D and 3D systems, a subset of workloads continue to benefit noticeably from increased \omct{sizes of these structures}.}

Increasing queue depths is not universally beneficial. In particular, increasing the \nmg{issue queue's} depth increases the \nmg{number of cycles spent in the issue stage}, which can negatively impact the performance of \emph{speculation-bound} workloads. Our analysis shows up to a 9.4\% slowdown (for the speculation-bound workloads) due to the larger pipeline fill time and pipeline bubbles on branch mispredictions. However, as we show in \sect{\ref{sec:revamp-memory}}, we can use the M3D-based system's main memory to reduce this overhead for speculation-bound workloads.

\subsubsection{Impact of Pipeline Frontend and Branch Prediction Units.}
\label{sec:arch-spec}
\begin{figure}[h] %
  \centering
  \begin{minipage}[b]{0.55\textwidth}
    \centering
    \includegraphics[width=\textwidth]{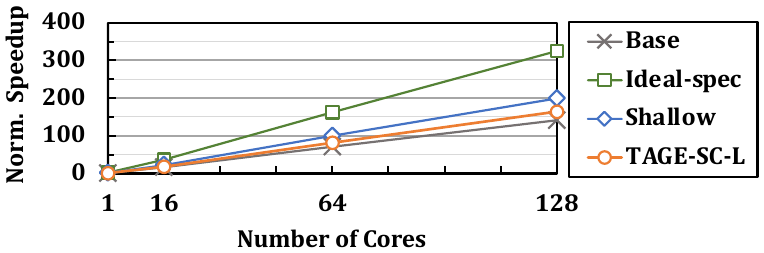}
    \caption{Effect of speculation and frontend overheads \omct{for a representative speculation-bound workload (\textsf{Triangle})}.}
    \label{fig:spec-frontend}
  \end{minipage}
  \hfill
  \begin{minipage}[b]{0.4\textwidth}
    \Copy{R3/1d-fig}{
    \centering
    \includegraphics[width=\textwidth]{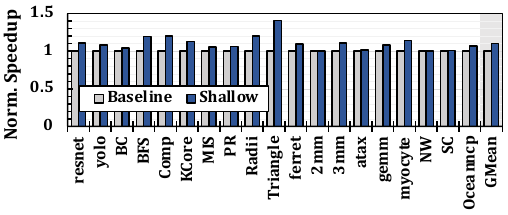}
    \caption{Performance benefits of an M3D system with a shallow pipeline \omct{across our evaluated workloads}.}
    }
    \label{fig:r3-branch}
  \end{minipage}
  \Description{}
\end{figure}

Due to the increased impact of the frontend and speculation bottlenecks in M3D, we further study their implications on performance. For a representative  workload \omct{whose performance is bottlenecked by branch misprediction rate (i.e., a speculation-bound workload)}, \texttt{Triangle}, \fig{\ref{fig:spec-frontend}} shows the performance of 
the baseline M3D-based system (\texttt{Base}) compared to  
M3D-based systems with  
\romnum{i}~a state-of-the-art branch predictor (\texttt{TAGE-SC-L})~\cite{seznec2016tage}, 
\romnum{ii}~an idealized frontend where the issue and dispatch structures in the pipeline take only one cycle (\texttt{Shallow}, i.e., a shallow pipeline that reduces the number of pipeline bubbles and fill latency in the event of branch misprediction), 
and \romnum{iii}~an \emph{idealized} branch prediction scheme, which leads to zero branch misprediction (\texttt{Ideal-spec}). 
First, we observe that \nmg{while \texttt{Ideal-spec} leads to a 2.3$\times$ speedup compared to the baseline M3D-based system,} using TAGE-SC-L~\cite{seznec2016tage} provides only a 14\% speedup. This is due to the large number of hard-to-predict branches in this workload. 
Second,  by reducing the overhead of branch misprediction, \texttt{Shallow} leads to up to 41\% speedup for the speculation-bound workload. \fig{\ref{fig:r3-branch}} demonstrates the speedup of \texttt{Shallow} over the baseline 64-core M3D system across all our evaluated workloads, showing 10\% average (40\% maximum) speedup.
\sect{\ref{sec:revamp-memory}} shows how \proposal leverages the high-bandwidth M3D main memory to alleviate the misprediction overhead and frontend bottlenecks in the M3D-based systems with varying memory latency values.

\noindent\ch{lavendergray}{\textbf{Takeaway 9.}}
\emph{Improving the processor core's frontend can lead to significant performance benefits by lowering the overhead of branch misprediction in the M3D \omct{system}.}

\subsubsection{Inter-Thread Communication.}
\label{sec:sync-opt}

\begin{figure}[b] %
  \centering
  \begin{minipage}[b]{0.48\textwidth}
  \centering
  \Copy{R2/5-fig}{
    \includegraphics[width=\textwidth]{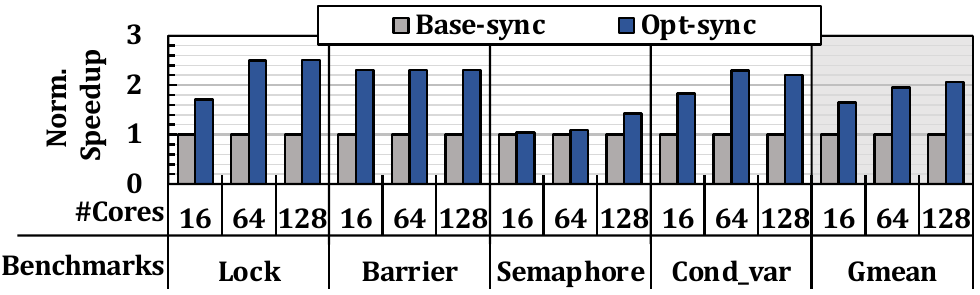}
    \vspace{0.25em}
    \caption{Speedup of the optimized synchronization for different synchronization primitives.}
    }
    \label{fig:rf-sync}
  \end{minipage}
  \hfill
  \begin{minipage}[b]{0.48\textwidth}
    \centering
    \includegraphics[width=\textwidth]{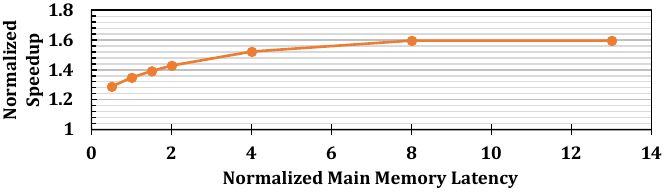}
    \vspace{-1em}
    \caption{Speedup of the optimized synchronization in M3D systems with different memory latency values \omct{for a synchronization-heavy workload (\textsf{Radii}).}}
    \label{fig:RF-sense-lat}
  \end{minipage}
\Description{}
\end{figure}

The large bandwidth of the M3D main memory can enable efficient feeding of data to a large number of threads. To take advantage of this high bandwidth, it is crucial to support efficient inter-thread communication. 
\fig{\ref{fig:rf-sync}} shows the speedup of four microbenchmarks~\cite{david2013everything,Libslock}, representing four commonly-used synchronization primitives in multi-threaded applications. We show the speedup of an M3D configuration where no overhead due to cache and memory latency is incurred during synchronization  (\texttt{Opt-sync}),\footnote{\raggedright We consider horizontal wire delay (which also exists in the baseline) and only reduce the overhead of accessing the cache and main memory for synchronization variables.} over an M3D configuration with baseline coherence-based synchronization (\texttt{Base-sync}). We observe on average 1.88$\times$ and up to 2.51$\times$ speedup.

We analyze the impact of synchronization on M3D systems with high main memory bandwidth but varying main memory latency. \fig{\ref{fig:RF-sense-lat}} shows the speedup of an M3D configuration where no overhead due to cache and memory latency is incurred during synchronization over an M3D configuration with baseline coherence-based synchronization for a synchronization-heavy workload (Radii). We observe that the synchronization overhead increases with larger memory latency.

\noindent\ch{lavendergray}{\textbf{Takeaway 10.}}
\emph{Reducing synchronization overhead can provide significant performance benefits for the M3D \omct{system}.}

\subsubsection{Latency of Non-Memory Micro-Operations.}
\label{sec:uop}
We model an M3D-based system where \emph{all} {\textmu}ops take one cycle and observe only a 5.4\% average speedup compared to the M3D baseline for compute-bound workloads since most functional units are pipelined.
We conclude that lowering the {\textmu}op latency does not lead to significant speedup in M3D.

\section{\proposal: Architecting \omf{Processor Core and Cache Hierarchy} Based on \omf{Implications of M3D Integration}}
\label{sec:case-study}

\begin{figure}[h]
\centering
\includegraphics[width=0.6\linewidth]{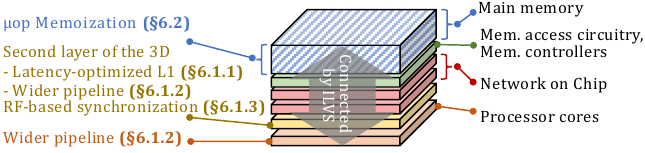}
\caption{\nmg{Overview of \omf{the key design decisions} in \proposal.}}
\label{fig:revamp-overview}
\Description{}
\end{figure}

We present \proposal, \omf{an optimized M3D-based system, based on the implications of M3D integration on cache hierarchy and processor core design.}
\omf{We target} performance bottlenecks in the M3D-based system
and \omf{take advantage of} opportunities offered by the M3D integration technology. \nmg{\fig{\ref{fig:revamp-overview}} shows an overview of \omf{the key design decisions}} in \proposal. \sect{\ref{sec:revamp-logic}} describes how \proposal  leverages the tight connectivity between M3D logic layers and
\sect{\ref{sec:revamp-memory}} shows how \proposal leverages M3D main memory to realize these \omf{design decisions}.

\subsection{Leveraging M3D Logic Layers}
\label{sec:revamp-logic}

\subsubsection{Cache Hierarchy.} 
As described in \sect{\ref{sec:arch-caches}}, even an optimized L2 cache (with increased size or reduced latency) does not significantly outperform the configuration with no L2 cache. Hence, we remove the L2 cache \omf{from the cache hierarchy} in \proposal, which leads to freeing up 32\% of the logic area. We then use this freed-up area for optimizations that alleviate the core bottlenecks and \omf{to widen and scale up various structures in the processor pipeline that enable greater instruction-level parallelism} (\sect{\ref{sec:revamp-width}}).
As described in \sect{\ref{sec:arch-caches}}, improving the L1 cache latency improves M3D performance due to the small AMAT in the M3D-based system.
The short ILVs of the M3D integration technology enable efficient division of the SRAM structures between two logic layers to reduce the overall wire lengths in the SRAM structures, leading to reduced latency~\cite{srinivasa1, srinivasa2, srinivasa2018monolithic, gopireddy2019designing}.
 \omf{Therefore, we leverage} an M3D layout~\cite{gopireddy2019designing}, which reduces L1 cache latency and planar footprint by 41\% and 44\%, respectively.

\subsubsection{Pipeline Width.}  
\label{sec:revamp-width}
To design wider pipelines, first, we increase the width \nmg{and depth} of the pipeline SRAM structures (e.g., L/SQ, ROB), leading to an \nmg{11.6}\% area overhead per core (\nmg{7.8}\% of the logic area). By exploiting the multi-layer SRAM structure supported by  M3D~\cite{srinivasa1,gopireddy2019designing}, we increase the size of these structures without increasing their latency and affecting the pipeline frequency.
Second, we increase the width of the decode structure and the functional units. 
Doubling the decode structure\grm{s} and functional units (including \grm{six} integer ALUs, \grm{one} FPU, and \grm{one} complex ALU) lead\grm{s} to \grm{a} 16.5\% area overhead \emph{per core} \mrevmic{(11.2\% of the logic area)}.\footnote{\nmg{We use McPAT~\cite{li2009mcpat} in 22nm technology for area analysis. Even though the absolute area of the logic components in our CNFET-based M3D baseline does not equal the values generated by McPAT, we are interested in the area \emph{ratio} of different components.}}  We exploit part of the extra die area freed up due to removing the L2 cache to double the \grm{number of} decode and execution units.

\subsubsection{Register File-Level Synchronization.} 
\label{sec:mech-rf}

To facilitate
high thread-level parallelism, we introduce a new fine-grained synchronization technique, using M3D's dense inter-layer connectivity, to improve inter-thread communication.
We use the thin ILVs between the logic layers in the M3D  system to perform fine-grained inter-thread communication in the register file (RF) instead of the memory hierarchy.
The thin ILVs enable increasing the bandwidth of the SRAM cells in the RF by adding extra access transistors to each cell. The short ILVs can efficiently connect the extra access transistors in a secondary logic layer to the SRAM cells located on the primary layer, without increasing the RF  latency. \proposal leverages this to add extra ports to the RF of each core and uses the extra ports for fine grained inter-thread communication. 
\begin{wrapfigure}{r}{0.3\textwidth}
\Copy{R1/2-fig}{
  \centering
  \includegraphics[width=\linewidth]{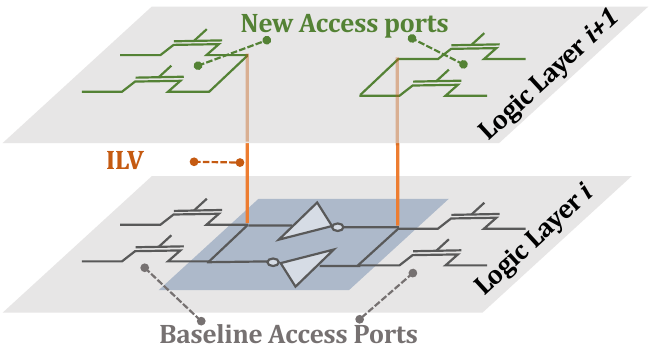}
  \caption{Vertical layout for extra ports.}
  }
  \label{fig:multi-port}
\end{wrapfigure}
\fig{\ref{fig:multi-port}} shows how fine-grained ILVs can add additional access ports to an SRAM cell. 
\emph{Small diameter} and \emph{vertical} connections are essential for adding extra ports efficiently. Due to the small ILV diameter, adding vertical ports increases the SRAM cell area at the 15nm technology node by less than 0.1\%~\cite{gopireddy2019designing}. However, for vertical connections with much larger diameters (e.g., 2.6$\mu$m TSVs), this overhead increases significantly~\cite{gopireddy2019designing}. For horizontal connections (i.e., in 2D), adding extra ports leads to not only area overhead, but also an increased lateral delay.

Due to RF's limited size, we leverage it \omct{only} for communications that involve small data, such as synchronization. 
While RF-level communication has been proposed in prior work~\cite{keckler1998exploiting}, the adoption of such proposals is challenging in 2D or 3D systems due to the limitations of adding extra ports to the RFs. 
Without adding extra ports, 
the synchronization operations need to compete with the pipeline's \omf{register operand accesses} for the limited RF ports.

\begin{figure}[h]
\Copy{R1/2-fig2}{
  \centering
  \includegraphics[width=0.9\linewidth]{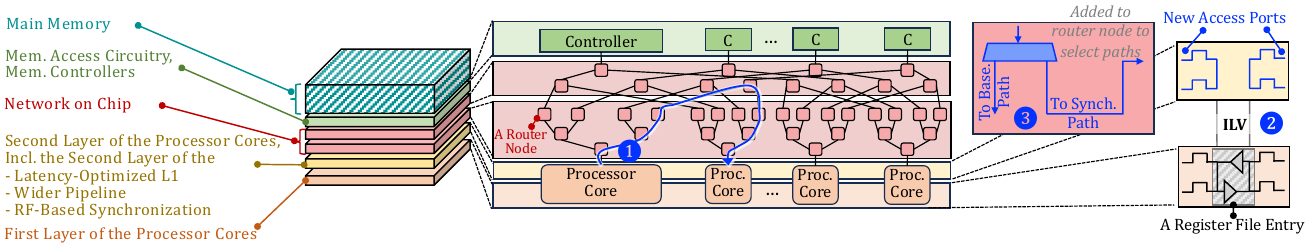}
  \caption{Overview of the register file-level synchronization.}
  }
  \label{fig:sync-details}
  \Description{}
\end{figure}

To enable fine-grained \omct{synchronization}, where one \omct{synchronization} variable is associated with a small amount of application data, each core writes the address of the data that is being \omct{synchronized} by the core to other cores' RF (instead of using \omct{synchronization} variables updated through the cache hierarchy and the coherence protocol).  \fig{\ref{fig:sync-details}} shows an overview of this process. When a core wants to send a \omct{synchronization} variable (\circled{1} in \fig{\ref{fig:sync-details}}), it can use the baseline interconnects (in this case, a meshes-of-trees network, same as the baseline M3D system), but instead of sending data to the memory hierarchy, it sends it to the register file of the other cores (\circled{2}). To this end, the router nodes need a simple selection structure (\circled{3}) to determine whether data needs to go through the baseline path to the memory hierarchy or to go through the synchronization path.

\begin{figure}[h]
  \centering
  \includegraphics[width=0.45\linewidth]{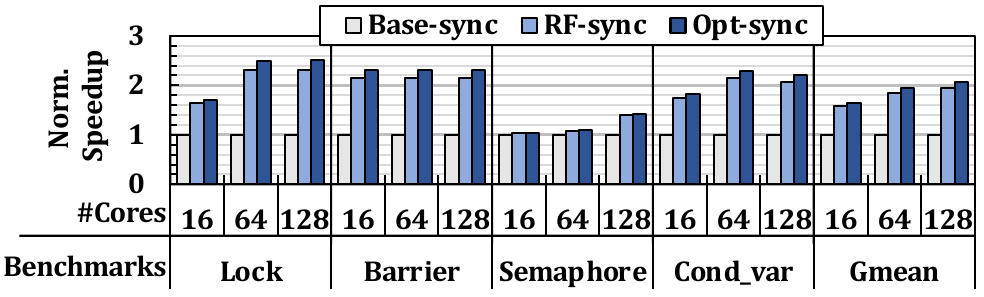}
    \vspace{-0.3em}
  \caption{Speedup of the register file-level synchronization \omct{ for different
synchronization primitives}.}
  \label{fig:rf-sync-opt}
  \Description{}
  \vspace{-0.1em}
\end{figure}

\nmg{\fig{\ref{fig:rf-sync-opt}} shows
the speedup of RF synchronization (\texttt{RF-sync}) compared to the baseline (\texttt{Base-sync}) for the synchronization microbenchmarks \omct{described in \sect{\ref{sec:sync-opt}}}.}
We observe on average 1.78$\times$ and up to 2.31$\times$ speedup \nmg{over the M3D-based system with the baseline coherence-based synchronization}.

\nmg{\head{Area and Timing Implications}} In fine-grained \omct{synchronization} schemes of real workloads, prior works~\cite{david2013everything,giannoula2021syncron} show that the number of synchronization variables that are active at any given time during the execution is small (typically up to 4 variables). Thus, there is no need for large additional space in the RF. 
RF takes 0.2\% of the logic area in our baseline. Therefore, the planar area overhead of adding extra ports to 4 entries of the RF is less than 0.001\% of the logic area.
Since the extra ports are connected to RF entries from another logic layer using the short ILVs, they do not change the structure and timing of the RF accesses for the baseline core operations. \omct{More details regarding security implications are in an extended version~\cite{ghiasi2022revamp3d}.}

\subsection{Leveraging M3D Memory Layers}
\label{sec:revamp-memory}

\nmg{\omf{We propose a new processor frontend design that}
memoizes the repetitive fetched, decoded, and re-ordered \uops, \omf{stores them in M3D main memory,} and turns off the relevant parts of the core pipeline when the same \uops execute in a loop. \omf{The high-bandwidth, energy-efficient M3D memory enables storing and loading the memoized instructions efficiently}. This addresses two key bottlenecks that we identify in the M3D baseline. 
First, it can alleviate the shifted M3D system's energy bottlenecks that are dominated by the processor core. 
As programmers commonly use loops in their codes, a significant number of instructions in programs are typically repeated\omct{~\cite{talpes2005execution,talpes2001power,solomon2003micro,conners1999compiler,fallin2014heterogeneous}}.
Therefore, turning off the frontend and reordering structures can lead to energy saving by eliminating \taco{a significant portion of the average energy per core}
in the baseline OoO core. 
Second, this technique alleviates the speculation bottleneck in the M3D system by reducing the pipeline fill latency and number of pipeline bubbles in the event of branch misprediction since the instructions do not need to be fetched, decoded, and reordered if they are already memoized.}

\taco{While prior works~~\cite{talpes2005execution,talpes2001power} have also explored \uop memoization with an Execution Cache (EC), these works leverage SRAM caches that incur large area overheads. However, \proposal leverages the high-bandwidth connection between logic and memory layers as a unique opportunity provided by M3D integration, and \omf{stores the memoized \uops} in the main memory.}
Prior work~\cite{talpes2005execution} shows that to fully benefit from \uop memoization, we need a large on-chip EC (e.g., up to 100KB) per core, which incurs high area overheads, occupying 15\% of the logic area of the baseline processor core. Considering a fixed area budget, this overhead can come at the cost of key core components and, therefore, sacrifice performance. Some prior works propose a smaller EC for area efficiency but at the cost of memoizability opportunities in applications with a large instruction footprint and large loops~\cite{talpes2005execution,talpes2001power, padmanabha2015dynamos,padmanabha2017mirage}. 
\taco{In contrast, using M3D main memory for storing the memoized \uops paves the way for architecting a large and scalable EC at low area cost. We use a small buffer to store the prefetched memoized \uops from main memory and provide a one-cycle access latency.}

\begin{figure}[b] %
  \centering
  
  \begin{minipage}[b]{0.48\textwidth}
    \centering
    \includegraphics[trim= 0mm 0mm 0mm 0mm,width=\linewidth]{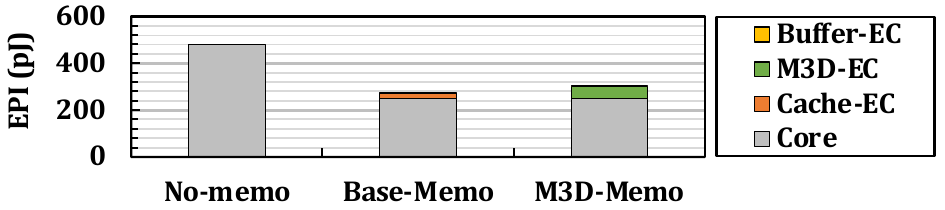}
    \caption{Energy per instruction of \omct{different $\mu$op memoization techniques}.}
    \label{fig:EPI}
  \end{minipage}
  \hfill
  \begin{minipage}[b]{0.48\textwidth}
    \centering
    \includegraphics[width=\textwidth]{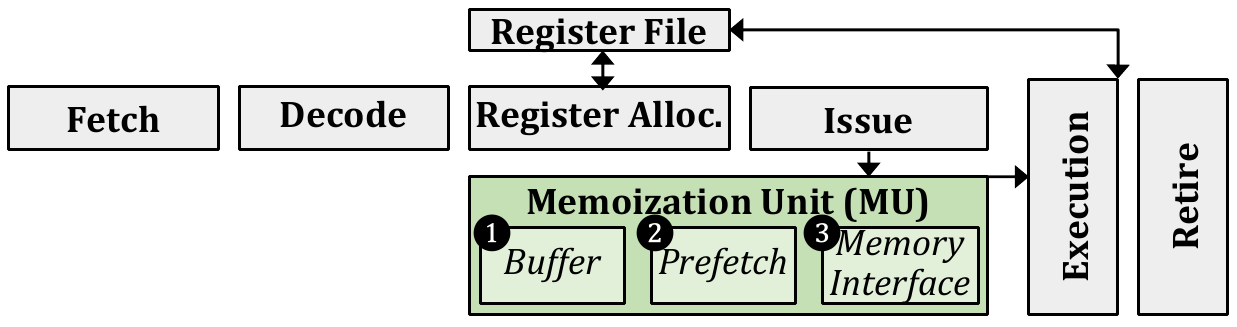}
    \caption{\taco{\proposal's memoization unit.}}
    \label{fig:mu}
  \end{minipage}
\Description{}
\end{figure}

\nmg{\fig{\ref{fig:EPI}} demonstrates} Energy per Instruction (EPI) in  (1) no memoization (\texttt{No-Memo}), (2) memoizing \uops in a 100KB-SRAM EC (\texttt{Baseline-Memo}), and (3) memoizing \uops in M3D main memory (\texttt{M3D-Memo}).  In (2) and (3), the frontend, decode, and reordering structures are power-gated when the memoized \uops execute. 
We break down the EPI between the core \omct{(\textsf{Core})}, the SRAM-based EC (\textsf{Cache-EC}) in \texttt{Baseline-Memo}, and the main memory-based EC (\textsf{M3D-EC}) and the small buffer (\textsf{Buffer-EC}) in \texttt{M3D-Memo}. 
We observe that \texttt{M3D-Memo} reduces EPI by 37\% over \texttt{No-Memo} \taco{and \omct{provides} comparable EPI compared to \texttt{Baseline-Memo}, while \omct{having} significantly lower area cost \omct{(i.e., 0.7\% of the core area used for \textsf{Buffer-EC} relative to 15\% of the core area used for \textsf{Cache-EC})}.}

\head{Implementation}  We implement \texttt{M3D-Memo} based on prior work~\cite{talpes2005execution} that proposes an SRAM-based EC.  \taco{As shown in \fig{\ref{fig:mu}}}, we place the \emph{Memoization Unit (MU)} after the issue stage. Note that the MU does not add an extra stage to the pipeline since instructions are issued to the execution stage and the MU in parallel. We use the high main memory bandwidth in M3D to memoize both the taken and not-taken paths of hard-to-predict branches. The MU has three main components: \circled{1}~a small 1.28KB SRAM buffer to provide one-cycle access latency to the memoized \uops, \circled{2}~a simple stride prefetcher that reads \uops from the main memory and fills the buffer to hide the memory access latency, and \circled{3}~a memory interface that uses two read/write ports for the taken and not-taken paths and an address bus. We calculate the area of the MU  using Synopsys Design Compiler and show that it occupies less than 5\% of the L1 cache area. We use a preserved address space to store the memoized \uops.

\head{Performance Implications}
\texttt{M3D-Memo} improves the performance \nmg{of speculation-bound workloads}  by reducing pipeline fill latency and bubbles.
However, \texttt{M3D-Memo} \emph{limits} renaming each architectural register to a specific register pool (similar to \cite{talpes2005execution}), which can negatively affect performance. 
Overall, the effect of reducing pipeline depth and fill latency in the M3D-based system outweighs this overhead, leading to up to 35.5\% speedup for speculation-bound workloads and a modest average speedup of 1.4\% across all \omf{evaluated} workloads.

\begin{figure}[h]
        \centering
        \vspace{-2pt}
        \includegraphics[width=0.4\linewidth]{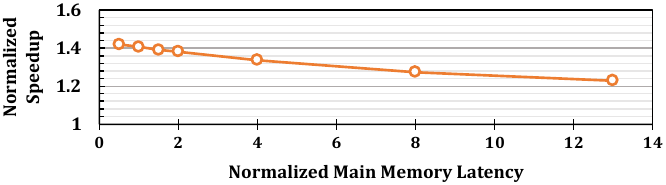}
                \vspace{-3pt}
        \caption{Performance impact of the \uop memoization technique on M3D-based systems with different memory latency values for \omct{a speculation-bound workload (\textsf{Tri}).}}
        \label{fig:memo-sense-lat}
        \Description{}
        \vspace{-5pt}
\end{figure}

\taco{To demonstrate the impact of this \omf{design} as M3D memory latency changes, \fig{\ref{fig:memo-sense-lat}} shows the performance of a 64-core M3D system with main memory-based \uop memoization for a speculation-bound workload (\texttt{Tri}). Performance at each point is normalized to the baseline M3D system with that memory latency. In all configurations, the M3D system has the same \omct{main memory} bandwidth. We find that the main memory-based \uop memoization achieves significant speedups (22–42\%) across a range of memory latencies, with  larger gains at lower \omct{main memory} latency values.}

\section{End-to-End Evaluation} 
\label{sec:revamp-eval}

\subsection{\nmg{End-to-End Speedup}}
\label{sec:revamp-perf-analysis}
\vspace{-0.5em}

\begin{figure*}[b]
\Copy{R2/1-fig}{
        \centering
       \includegraphics[trim= 0mm 0mm 0mm 0mm,width=0.9\linewidth]{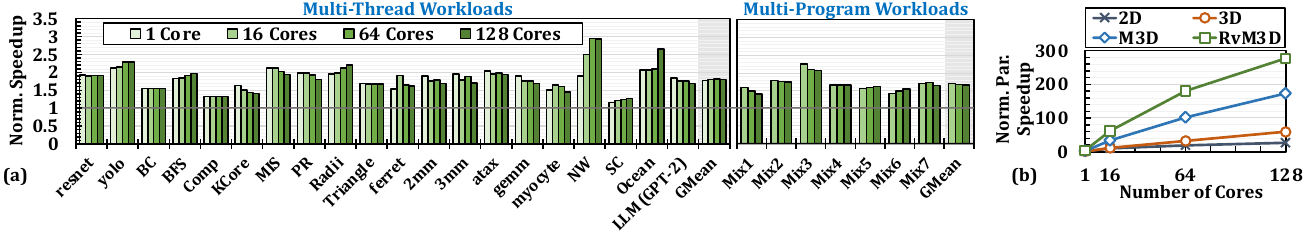}
        \vspace{-0.2em}
        \caption{\proposal's Speedup over (a) the M3D baseline, and (b) the 2D and 3D baselines on average, across all evaluated workloads.}
        }
        \label{fig:E2E-all}
        \Description{}
        \vspace{-0.2em}
\end{figure*}

\fig{\ref{fig:E2E-all}}\textit{(a)} shows the speedup of \proposal (with all \omf{design decisions} of \sect{\ref{sec:case-study}}) over the state-of-the-art M3D system, across all multi-threaded workloads and multi-programmed workload mixes. Each data point shows the speedup of \omct{\proposal} with $N$ cores over the baseline M3D with $N$ cores, where $N\in\{1, 16, 64, 128\}$.
We make two observations. First, we observe a significant performance improvement, on average \nmg{79.4\% and 67.6\%} speedups \nmg{across all core counts} for \nmg{multi-threaded workloads and multi-programmed workload mixes}, respectively. Second, the \omct{proposed} \omf{design decisions} improve the performance of \emph{all \omf{evaluated} workloads} (i.e., no workload \omct{loses performance over the baseline M3D system}).

\fig{\ref{fig:E2E-all}}\textit{(b)} shows \proposal's speedup, \nmg{averaged across all \omf{evaluated} workloads, over} the state-of-the-art 2D, 3D, and M3D-based systems. We observe that \proposal provides
significant benefits, resulting in \nmicro{7.14$\times$ and 4.96$\times$} speedups over the 2D and 3D systems, respectively, further increasing the  performance benefits of M3D integration.

\begin{figure}[b] %
  \centering
  
\begin{minipage}[b]{0.48\textwidth}
\centering
\includegraphics[width=\linewidth]{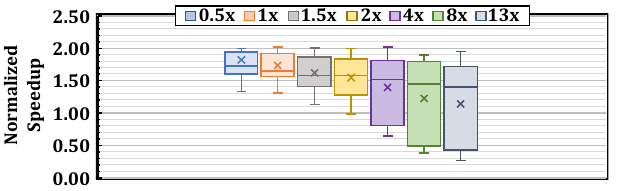}
\caption{\proposal's speedup with various memory latencies \omct{across all evaluated workloads. Main memory latency values are normalized to the baseline M3D memory latency of 5 ns}.}
\label{fig:sense-e2e}
\end{minipage}
  \hfill
\begin{minipage}[b]{0.48\textwidth}
\centering
\includegraphics[width=\linewidth]{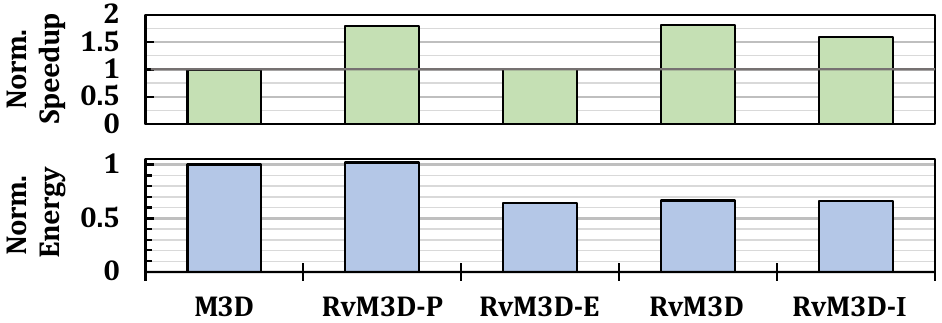}
\vspace{0.45em}
\caption{Speedup and energy of \proposal's configurations, \omct{averaged across all evaluated workloads}.}
\label{fig:perf-energy}
\end{minipage}
\Description{}
\end{figure}

\taco{For varying main memory latency values, \fig{\ref{fig:sense-e2e}} shows the distribution of the end-to-end speedup of \proposal
over the baseline M3D-based system across all the workloads. \proposal's speedup at each memory latency point is reported relative to the performance of the baseline system at that same latency. Both \proposal and baseline feature 64 cores. We observe that \proposal provides speedup for all \omf{evaluated} workloads even when memory latency increases \omct{by} 2$\times$, while some workloads exhibit slowdowns at higher latencies \omct{(e.g., 4$\times$-13$\times$ higher than the baseline M3D main memory latency)}. This is because, as shown in \fig{\ref{fig:cache-sense-lat}}, some design \omf{decisions} in \proposal are primarily tailored for smaller latency values.
While not all \proposal design choices lead to speedup for M3D-based systems with high latency, high-bandwidth connectivity between logic and memory layers in the M3D-based system can still provide several benefits (e.g., RF-based inter-thread synchronization and efficient \uop memoization), which provide benefits for M3D-based systems with different memory latency values. Therefore, depending on the configuration of the M3D-based system and its main memory characteristics, different subsets of \proposal's \omf{design decisions} can be adopted to enhance both performance and energy efficiency.}

\subsection{\nmg{Analysis of Different \proposal Configurations}}
\label{sec:case-study-energy}

\nmg{We evaluate the benefits of four different \proposal configurations, each optimized for different goals. \fig{\ref{fig:perf-energy}} shows the normalized speedup and energy consumption of each configuration relative to the baseline M3D system, averaged across all \omf{evaluated} workloads. First, \texttt{RvM3D-P} represents the \proposal configuration that \omct{uses only} the performance-improving \omf{design decisions} in \sect{\ref{sec:revamp-logic}}. Second, \texttt{RvM3D-E} represents the configuration that \omct{uses only} the \omf{design decisions} in \sect{\ref{sec:revamp-memory}}, to reduce energy. Third, \texttt{RvM3D} combines both sets of \omf{design decisions}.
Fourth, \texttt{RvM3D-I} represents a low-frequency implementation of \proposal that is iso-power with the M3D baseline.}  We make four observations.
First, \nmg{\texttt{RvM3D-P}} leads to similar average energy consumption compared to the M3D baseline (with only a modest 2\% increase). This is because, given the high energy efficiency of the M3D main memory, removing the L2 does not significantly increase energy.  
Furthermore, increasing the width of the pipeline SRAM structures using the M3D vertical layout~\cite{gopireddy2019designing} does not increase their energy per access since their overall planar wire length does not increase. 
Second, \nmg{\texttt{RvM3D-E}} leads to 36.\nmg{6}\% lower average \nmg{energy} consumption since, for each of our workloads, we observe a high degree of memoizability, where at least 99\% of their dynamic instructions follow the same schedule.
\texttt{RvM3D-E} leads to an average speedup of 1.4\%. 
Third,
\nmg{\texttt{RvM3D} leads to} a \nmg{79.4}\% average speedup, \omct{while also reducing energy consumption by 33.3\%}. Although the performance-improving \omf{design decisions} lead to an \nmg{84}\% power increase over the baseline M3D, the memoization technique allows this improvement at only \nmg{18}\% higher power consumption, \omct{which, combined with the large speedup, results in energy reduction}. 
Fourth, \nmg{\texttt{RvM3D-I}} outperforms the 4GHz baseline M3D by \nmg{60.5}\%, without an increase in power consumption. We conclude that our approach increases the M3D system's performance per Watt by both improving performance and lowering \nmg{energy} consumption.

\subsection{Overall Area Reduction}

We summarize the area impact of \proposal in Table~\ref{table:area}. We observe an \emph{overall} \nmg{12.3}\% area reduction compared to the baseline M3D-based system. The freed-up area can be used to further improve performance (e.g., by adding more cores).
\nmg{\proposal re-purposes the logic layer used for the L2 cache in the baseline M3D-based system (shown in \fig{\ref{fig:background-n3xt}}) to add all the needed components (e.g., the extra RF ports, the second layer of the L1 cache, and the second layer of the wider pipeline's SRAM structures and its additional functional units), while consuming smaller logic area than the L2 cache.}

 \begin{table}[h]
 \scriptsize
 \centering

\caption{Area Impact of \proposal.}
\vspace{-0.3em}
 \label{table:area}
\resizebox{0.45\columnwidth}{!}{%
\begin{tabular}{cc}
\toprule
\textbf{Proposed Change}   & \textbf{Area Impact (\% of logic area)}\\
\midrule
L2 cache \omct{(256 KB per core)} removal      &     $-32\%$   \\
Wider pipeline  &     $+\nmg{19}\%$  \\
EC Buffer       &     $+0.7\%$   \\
Register file overheads & $< +0.01\%$ \\
\midrule
Total           &     -\nmg{12.3}\%\\
\bottomrule
\vspace{-2em}
\end{tabular}}
 \end{table}

\section{Discussion}
\label{sec:discussion}

\head{Multiple M3D Stacks}
While a system with a single M3D chip \omct{is projected to} provide hundreds of gigabytes of main memory~\cite{aly2018n3xt, shulaker2017three}, for even larger requirements, a system can employ multiple M3D chips. In that case, off-chip access costs more overhead than on-chip access, and techniques that alleviate the remote access overhead using scheduling, data placement, replication, or migration can be orthogonally applied\omct{~\cite{ahn2015scalable, tang2017data,chen2022pimcloud,pattnaik2016scheduling}}.
Advances in integration technology can further reduce the cost of off-chip accesses by enabling the monolithic integration of \emph{multiple} M3D stacks~\cite{radway2021future}.

\head{Hardware/Software Co-Design} 
Memory bottleneck affects both hardware and software in today’s systems. This work explores the implications of M3D on the hardware design of cores and caches. We believe that it is essential to also revisit the software stack to better leverage the underlying M3D hardware. In  operating systems, memory management and process management units can be revisited to be aware of M3D's trade-offs.
In our work, we considered 2MB page sizes. Based on our analysis with Virtuoso\omct{~\cite{kanellopoulos2025virtuoso,virtuosogit}}, a state-of-the-art framework for the software and hardware components of the virtual memory subsystem, we observe that in our M3D baseline, using \omct{a} 2MB page size leads to 5.7\% average speedup across our evaluated workloads compared to using \omct{a} 4KB page size. We believe further detailed analysis of the system software stack \omct{and new virtual memory ideas~\cite{hajinazar2020virtual,kanellopoulos2025virtuoso,kanellopoulos2023victima,kanellopoulos2023utopia,kanellopoulos2026revelator}} in the context of the M3D technology would be an important direction to explore, and 
we hope our analysis of new bottlenecks and opportunities of M3D systems guides future hardware/software co-design efforts for this technology.

\section{Related Work} 
\label{sec_related}

\taco{To our knowledge, this is the first work to \omf{re-examine} the processor core and cache hierarchy based on rigorous bottleneck analysis of a wide range of workloads and the new trade-offs imposed by M3D integration.}

\head{Device and \mrev{Circuit}-\mrev{L}evel  \mrev{I}ntegration} Many works (e.g.,~\cite{ku2016physical, liu2012design, panth2014design, chang2017design,samal2014fast, shukla2019overview, lee2018performance, samal2014full, brocard2017transistor,vemuri2023efficient,prakash2023monolithic,vemuri2023fdsoi,lu20242d,xie2023monolithic,guan2023monolithic,pendurthi2024monolithic}) focus on the design of efficient M3D devices with respect to performance, power, energy, area, temperature, yield and reliability. 
Other studies show the benefits of M3D technology in comparison to 3D-stacked  technologies~\cite{samal2016monolithic, nayak2015power}.
Several works (e.g.,~\cite{jagasivamani2019analyzing, walden2021monolithically, jagasivamani2019design}) study M3D physical integration and layout. Several works explore heterogeneous device layers in M3D (e.g.,~\cite{zokaee2019magma, murali2020heterogeneous, chatterjee2021power, murali2021heterogeneous,du2023monolithic}).

\head{Designing M3D Processors}  
Recent works propose 3D partitioning techniques for logic-only M3D processors (e.g.,~\cite{stow2019efficient,joardar20213d++,gopireddy2019designing}) to improve the latency of different components by decreasing their wire lengths.
Other works show how to leverage M3D to design efficient SRAMs (e.g.,~\cite{srinivasa2018monolithic,srinivasa1,srinivasa2,kong2017architecting})  or  NoCs (e.g.,~\cite{jagasivamani2018memory,stow2019efficient}). Some works (e.g.,~\cite{jagasivamani2018memory, walden2021monolithically}) discuss the memory system challenges, such as optimizing the memory array structures and developing techniques to increase memory parallelism (e.g., SIMD units and non-blocking loads). 
\taco{Some works explore the benefits of M3D integration of non-volatile memory on top of last-level caches and how to design the interface between them (e.g.,~\cite{walden2021monolithically,jagasivamani2020tileable,jagasivamani2019analyzing}). }
While prior works propose various optimizations, we analyze the implications of M3D on \romnum{i}~the real-world application bottlenecks and \romnum{ii}~architectural designs that have been conventionally specialized to tackle the memory bottleneck. For example, while prior works propose optimizations to caches in M3D~\cite{gopireddy2019designing, aly2018n3xt}, we show that such optimizations are \emph{not essential} beyond L1 for M3D-based systems with low memory latency values.

\head{Accelerating Specific Applications}
\omf{Various} works (e.g.,~\cite{srimani2023ultra,chen2021marvel,felfldate2020,yu2020spring, ko2020efficient, huang2019ragra, hanif2020resistive, joardar20213d++, wu2018hyperdimensional, wu2018brain,kang2023monolithic,kumar2023frontiers}) design domain-specific accelerators in M3D.
Some works (e.g.,~\cite{yu2018monolithic}) propose M3D FPGAs. \nmg{While our work focuses on general-purpose systems, 
\revtaco{we hope that}
our insights can also inspire domain-specific systems.}

\taco{\head{Processing-Using-Memory in M3D-Based Systems} Some works (e.g.,~\cite{li2023spatial,zhang20233d,du2023monolithic}) design processing-using-memory systems specialized for M3D-integrated systems.}

\head{3D Caches} Prior works propose 3D caches to achieve low latency and high bandwidth, such as AMD's 3D SRAM V-Cache~\cite{vcache} and SILO~\cite{SILO}.
However, due to their limited capacity,
\revtaco{they} cannot fully address main memory bottlenecks.

\omct{\head{3D-Stacked Memory} Various works (e.g.,~\cite{ahn2015scalable,drumond2017mondrian,boroumand2019conda,mutlu2022modern,NDC_ISPASS_2014,singh2019napel,azarkhish2016logic,azarkhish2018neurostream,top-pim,RVU,mutlu2019processing,NIM,gao2017tetris,mutlu2019enabling,hsieh2016transparent,cali2020genasm,boroumand2021mitigating,boroumand2021google,boroumand2021polynesia,fernandez2020natsa,LiM_3D_FFT_MM,akin2014hamlet,gao2016hrl,farmahini2015nda,boroumand2018google,nai2017graphpim,kim2018grim,PEI,kwon202125,lee2021hardware,ghiasi2022alp,mutlu2025memory,hsieh2016accelerating,nair2015active,oliveira2022accelerating,singh2021fpga}) propose architectural optimizations for systems with 3D-stacked memory. However, these techniques do not target new properties of M3D-based systems.}

\omct{\head{Lean Core Designs} Several works (e.g.,~\cite{fallin2014heterogeneous,padmanabha2017mirage,lukefahr2012composite,padmanabha2015dynamos}) propose lean core designs to reduce area and/or energy consumption. However, these techniques do not target the unique properties of M3D technology. As shown in \sect{\ref{sec:revamp-memory}}, some of these techniques (e.g., $\mu$op memoization) can be enabled and adopted more efficiently by leveraging the unique properties of the M3D technology.}

\section{Conclusion}
\vspace{-0.3em}

\taco{In this work, we \omct{rethink and redesign} the processor core and cache hierarchy, given the fundamentally new trade-offs of M3D. We show that for a variety of workloads on a state-of-the-art M3D system, the performance and energy bottlenecks \omf{substantially} shift from the main memory to the processor core and cache hierarchy, emphasizing the need to revisit current designs. We conduct a rigorous design space exploration of the key components of the processor core and cache hierarchy to understand the implications of the M3D system’s shifted bottlenecks on their design. 
Based on these implications and by \omf{taking advantage of} the new opportunities of M3D integration, we design a new M3D system, \proposal. We show that \proposal provides significant performance improvement and energy reduction while achieving a smaller logic area compared to the state-of-the-art M3D system. \omct{We hope that the analysis, insights, and design directions presented in this work inspire future research on rethinking system designs based on the unique opportunities of monolithic 3D integration.}}

\omct{\section*{Acknowledgments}
We thank the anonymous reviewers of TACO 2026 for their feedback.
Earlier versions of this work were submitted to SIGMETRICS 2021, ISCA 2022, MICRO 2022, and ASPLOS 2023, and unfortunately, met some not-so-receptive feedback while also receiving some constructive feedback.
We thank the SAFARI group members for feedback and the stimulating, inclusive, intellectual, and scientific environment. We acknowledge the generous gifts from our industrial partners, including ASML, Google, Huawei, Intel, and Microsoft. This work, along with our broader work in Processing-in-Memory and memory systems, is supported in part by the Semiconductor Research Corporation (SRC), the ETH Future Computing Laboratory (EFCL), AI Chip Center for Emerging Smart Systems (ACCESS), a Google Security and Privacy Research Award, and the Microsoft Swiss Joint Research Center.}

\bibliographystyle{ACM-Reference-Format}
\bibliography{refs}

\end{document}